\documentclass[10pt,a4paper]{article}
\usepackage[T1]{fontenc}
\usepackage[utf8]{inputenc}
\usepackage{lmodern,amsmath,amssymb,amsthm,bm,mathtools,microtype,mathrsfs}
\usepackage[margin=20mm]{geometry}
\usepackage{graphicx,booktabs,array,enumitem}
\usepackage[hidelinks]{hyperref}
\hypersetup{pdftitle={Dynamical obstructions to primary Proca hair and its logarithmic Maxwell continuation},pdfkeywords={primary hair, generalized Proca, angular instability, Maxwell deformation, black hole stability}}
\newcommand{\dd}{\mathrm d}
\newcommand{\mpl}{M_{\mathrm{Pl}}}

\newcommand{\Kt}{\mathcal K_1}
\newcommand{\doi}[1]{\href{https://doi.org/#1}{\nolinkurl{#1}}}
\newtheorem{proposition}{Proposition}
\setlist{itemsep=2pt,topsep=3pt}

\title{Dynamical obstructions to primary Proca hair and its logarithmic Maxwell continuation}
\author{\begin{minipage}{\dimexpr\textwidth-2\tabcolsep\relax}\centering
Anirudh Pradhan$^{1}$, K. Ghaderi$^{2,*}$, Ulug'bek Eshkarayev$^{3}$, \\
Sardor Sabirov$^{4}$, and Javlon Rayimbaev$^{5,6,7}$ \\[0.1em]
\footnotesize $^{1}$Centre for Cosmology, Astrophysics and Space Science (CCASS), GLA University, Mathura 281406, U.P., India\\
\footnotesize $^{2}$Department of Physics, Mari.C., Islamic Azad University, Marivan, Iran\\
\footnotesize $^{3}$Department of Pedagogy and Psychology, Termez University of Economics and Service, Termez, Uzbekistan\\
\footnotesize $^{4}$Department of General Professional Sciences, Mamun University, Khiva, Uzbekistan\\
\footnotesize $^{5}$Harbin Institute of Technology, Harbin 150001, People's Republic of China\\
\footnotesize $^{6}$University of Tashkent for Applied Sciences, Str. Gavhar 1, Tashkent 100149, Uzbekistan\\
\footnotesize $^{7}$Institute of Theoretical Physics, National University of Uzbekistan, Tashkent 100174, Uzbekistan\\[0.2em]
\footnotesize $^{*}$Corresponding author: \texttt{k.ghaderi@iau.ac.ir}
\end{minipage}}
\begin{document}
\setlength{\abovedisplayskip}{8pt plus 2pt minus 3pt}
\setlength{\belowdisplayskip}{8pt plus 2pt minus 3pt}
\setlength{\belowdisplayshortskip}{4pt plus 2pt minus 2pt}
\maketitle\vspace{-10mm}
\begin{abstract}
We examine the dynamical admissibility of primary black hole hair in a regularized vector theory of Gauss--Bonnet gravity. The original nontrivial constant norm family cannot satisfy both positive dipole kinetic energy and real nondegenerate radial tensor characteristics. A regular Hamiltonian reduction treats the degenerate boundaries. For asymptotically flat exteriors on the nondegenerate radial branch, adding a Maxwell term for the same vector requires a varying norm. Its positive logarithmic continuation has finite mass at spatial infinity and numerically constructed regular future horizons. The constructed backgrounds satisfy the radial tensor and transverse vector conditions at the sampled points. However, the constrained even angular system has a longitudinal mode with negative squared angular speed. We derive its inverse logarithm coefficient, including both the vector constraint and quartic interaction, and verify it using independent gauge reductions and the full backgrounds. This local angular obstruction persists despite the compatible radial conditions.
\end{abstract}

\section{Introduction}
An exact black hole geometry is physically viable only if its perturbations have consistent dynamics. In generalized Proca theory, different interactions govern the longitudinal and transverse vector modes. Horizon regularity and second order field equations alone do not determine their kinetic signs or characteristics \cite{Heisenberg2014,Kase2018}.

The regularized vector theory of Ref.~\cite{Charmousis2025} admits an asymptotically flat family with an independent metric charge and a vector constant that changes the norm while preserving the geometry. Related constructions include rotating solutions and higher Lovelock interactions \cite{Charmousis2026,Hassaine2026,FernandesRotating2026,FernandesTower2026}. Equilibrium thermodynamics has also been examined for a constrained rotating subfamily \cite{Liu2026}. Optical signatures and the ringing of scalar and Dirac test fields probe the resulting geometries \cite{KonoplyaOptics2026,KonoplyaEcho2026}. Their dynamical admissibility also requires analysis of the coupled metric and vector perturbations. We ask whether the norm parameter can make the original family admissible and whether a kinetic term for the same vector can remove its obstruction.

Our first result is an action specific incompatibility for the constant norm family. Direct dipole and helicity two expansions show that positive dipole kinetic energy and real nondegenerate radial tensor characteristics cannot coexist. At the degenerate boundaries, retaining the rotational constraints identifies the surviving physical canonical pair, while the tensor principal equation has a repeated characteristic. The general odd perturbation formalism \cite{Kase2018} and instabilities of other vector actions \cite{Aoki2023} provide the established context for this specialization.

Black holes with vector kinetic terms in broader generalized Proca theories are known \cite{Heisenberg2017}. An obstruction to adding a Maxwell term also exists for aligned null Kerr--Schild configurations in five dimensions \cite{Hassaine2026}. We allow both metric functions and the norm to vary on a static four dimensional branch. For the fixed interactions considered here, asymptotic flatness excludes nontrivial hair on the nondegenerate constant norm radial branch. The varying norm response instead leads to a singular far problem with a positive logarithmic branch. An exact ingoing reduction constructs finite deformation exteriors with regular future horizons and finite ADM mass. We check these solutions with independent integrations and all four background equations. Their radial tensor and transverse vector forms satisfy the common frame criterion of Ref.~\cite{Babichev2018} at the sampled exterior points.

Angular instabilities of cubic vector Galileon black holes are established \cite{DeFelice2024}. Our second dynamical result concerns the present logarithmic continuation, where the first angular term cancels. Both the Maxwell constraint and quartic interaction enter the next coefficient, giving a negative longitudinal squared angular speed with inverse logarithm decay. We derive all three physical even angular principal modes, recover the published cubic limit, and verify the physical roots in independent gauges. The main contribution is the first nonzero angular coefficient on this logarithmic branch, obtained after imposing the vector and metric constraints. The exact background reduction and axial classification establish how this branch relates to the original family. All conclusions concern local admissibility conditions on the stated branches.

Sections~\ref{sec:theory}--\ref{sec:tensor} establish the axial incompatibility; Section~\ref{sec:boundaries} treats its endpoints. Sections~\ref{sec:maxwell}--\ref{sec:far} derive the deformation and far balance. Sections~\ref{sec:finite} and \ref{sec:angular} construct the backgrounds and derive the angular obstruction. Section~\ref{sec:conclusion} gives the physical classification. The appendices contain variation, convention, angular constraint and boundary Hamiltonian details.

\section{Theory and exact background}\label{sec:theory}
\subsection{Action, conventions, and vector equation}
We use signature $(-,+,+,+)$ and curvature convention
\begin{equation}
 R^\rho{}_{\sigma\mu\nu}=\partial_\mu\Gamma^\rho_{\nu\sigma}
 -\partial_\nu\Gamma^\rho_{\mu\sigma}
 +\Gamma^\rho_{\mu\lambda}\Gamma^\lambda_{\nu\sigma}
 -\Gamma^\rho_{\nu\lambda}\Gamma^\lambda_{\mu\sigma}.
\end{equation}
The action is \cite{Charmousis2025}
\begin{align}
 I&=\frac{\mpl^2}{2}\int\dd^4x\sqrt{-g}\,[R-\alpha\mathcal L_V],\label{eq:action}\\
 \mathcal L_V&=4G^{\mu\nu}W_\mu W_\nu
      +8W^2\nabla_\mu W^\mu+6(W^2)^2,\qquad W^2=W_\mu W^\mu.\label{eq:LV}
\end{align}
Here $\mpl^2>0$, $[\alpha]=L^2$, and the orthonormal components of $W_\mu$ have dimension $L^{-1}$. We define $X=-W^2/2$ and use $f$ for the radial metric function. There is no term proportional to $W_{\mu\nu}W^{\mu\nu}$, where $W_{\mu\nu}=2\nabla_{[\mu}W_{\nu]}$.

For $\alpha\ne0$, variation with respect to the covariant vector gives
\begin{equation}
 \mathcal E_W^\mu\equiv G^{\mu\nu}W_\nu
 +2W^\mu\nabla_\nu W^\nu-\nabla^\mu W^2+3W^2W^\mu=0.\label{eq:vectorE}
\end{equation}
The derivative of $W^2$ is retained before imposing a constant norm. The cubic interaction in Eq.~\eqref{eq:LV} contributes only first derivatives to this equation, but it is essential to the background relation.

With the explicit convention $\mathcal L_3=G_3\nabla_\mu W^\mu$ and
\begin{equation}
 \mathcal L_4=G_4R+G_{4,X}\big[(\nabla_\mu W^\mu)^2
              -\nabla_\mu W_\nu\nabla^\nu W^\mu\big],
\end{equation}
the action, apart from its positive overall factor, is represented by
\begin{equation}
 G_2=-24\alpha X^2,\qquad G_3=16\alpha X,\qquad G_4=1-4\alpha X.\label{eq:dictionary}
\end{equation}
This dictionary fixes the sign convention for $\mathcal L_3$. It follows directly from Eq.~\eqref{eq:action} and the integration identity in Appendix~\ref{app:dictionary}.

\subsection{Independent spherical variation}
Before any relation among the vector components is imposed, take
\begin{equation}
 \dd s^2=-N^2 f\dd t^2+\frac{\dd r^2}{f}+r^2\dd\Omega^2,
 \qquad W=a(r)\dd t+b(r)\dd r.\label{eq:ansatz}
\end{equation}
After removal of the Einstein boundary derivative, the radial Lagrangian is
\begin{align}
 L_r&=2N(1-f-rf')-\alpha Nr^2\,[4\mathcal I+8Y\Theta+6Y^2],\label{eq:radialL}\\
 Y&=-\frac{a^2}{N^2f}+fb^2,\qquad
 \Theta=fb'+\left(f'+\frac{fN'}N+\frac{2f}r\right)b,\\
 \mathcal I&=-\frac{a^2}{N^2f}\frac{rf'+f-1}{r^2}
 +fb^2\left(\frac{rf'+f-1}{r^2}+\frac{2fN'}{rN}\right).
\end{align}
The omitted constant factor is positive. We vary $N,f,a,b$ independently. Substituting $Y=-2c$ into the action before variation would discard equations needed below.

Let $D=M-Q$ and write
\begin{equation}
 N=1,\quad a=A(r),\quad b=\frac{g(r)}f,\quad
 A^2=g^2+2cf,\quad g=u+cr,\quad f=1-\frac{2D}{r}-2ru.\label{eq:branchrelations}
\end{equation}
Substitution into the four Euler--Lagrange equations, after variation, gives
\begin{equation}
 E_f=E_a=E_b=0,\qquad
 E_N=4\frac{\dd}{\dd r}\big(r^2u+2\alpha r u^2\big).\label{eq:ELverification}
\end{equation}
Thus the remaining equation is the first integral
\begin{equation}
 r^2u+2\alpha r u^2=Q.\label{eq:firstintegral}
\end{equation}
The independent off diagonal metric equation is accounted for by the static shift identity described in Appendix~\ref{app:variation}. The angular metric equation follows from covariance once the radial and vector equations hold.

The asymptotically flat branch is conveniently written without a subtractive cancellation:
\begin{align}
 u(r)&=\frac{2Q}{r^2\left(1+\sqrt{1+8\alpha Q/r^3}\right)},\label{eq:u}\\
 f(r)&=1-\frac{2(M-Q)}r
 +\frac{r^2}{2\alpha}\left(1-\sqrt{1+\frac{8\alpha Q}{r^3}}\right).
 \label{eq:lapse}
\end{align}
These reproduce the family of Ref.~\cite{Charmousis2025}. The constant $c=X$ has dimension $L^{-2}$. It changes $A$ and $g$, but not $f$. Our local analysis requires a real vector, $g^2+2cf\ge0$, and a regular static interval with $f>0$. It does not assume that every value of $c$ defines an acceptable global vector boundary condition.

\subsection{Regularity and limiting cases}
For $c=0$, choose the future regular orientation $A=g=u$. With $v=t+\int\dd r/f$,
\begin{equation}
 \dd s^2=-f\dd v^2+2\dd v\dd r+r^2\dd\Omega^2,
 \qquad W=u\dd v.\label{eq:EF}
\end{equation}
The vector is null everywhere and regular at a simple horizon when $u$ is finite. The divergence of the static component $b=u/f$ is therefore a coordinate effect. For nonzero $c$, the regular local sign is $A\to g$ at the future horizon, and $(g-A)/f=-2c/(g+A)$ is finite when $g_h\ne0$.

At large radius on the null branch,
\begin{align}
 u&=\frac Q{r^2}-\frac{2\alpha Q^2}{r^5}+O(r^{-8}),\\
 f&=1-\frac{2M}r+\frac{4\alpha Q^2}{r^4}+O(r^{-7}).\label{eq:asymptotic}
\end{align}
The coefficient of $r^{-1}$ fixes the geometric Arnowitt--Deser--Misner (ADM) mass $M$, with $[M]=[Q]=L$. The metric approaches Schwarzschild for $Q\to0$. At $\alpha=0$ the vector interaction disappears from the action; it is not a limit with a finite, canonically normalized vector kinetic term.

\section{Axial dipole and its reduced kinetic form}\label{sec:dipole}
\subsection{Gauge structure and rotational constraint}
For an axisymmetric representative of the odd $\ell=1$ sector, introduce
\begin{align}
 \dd s^2={}&-f\dd t^2+\frac{\dd r^2}{f}+r^2\dd\theta^2
 +r^2\sin^2\theta\,[\dd\phi+\epsilon\Omega(t,r)\dd t
                         +\epsilon P(t,r)\dd r]^2,\\
 W={}&A\dd t+\frac gf\dd r+\epsilon a_1(t,r)\sin^2\theta\dd\phi.
 \label{eq:oddansatz}
\end{align}
The amplitude $a_1$ is dimensionless as a covariant angular component. The order $\epsilon^2$ metric terms are retained in the covariant expansion. The angular coordinate change shifts $\Omega$ and $P$ by derivatives of a function $\Lambda(t,r)$, leaving
\begin{equation}
 \mathcal F_1=\partial_tP-\partial_r\Omega
\end{equation}
invariant. The vector amplitude $a_1$ is invariant under this linear axial coordinate transformation because the background has no angular component.

The two rotational equations imply a quantity constant in both $t$ and $r$. It is the homogeneous angular momentum perturbation. We choose its zero value when discussing the propagating vector mode. This fixes an independent rotational sector; it does not impose $a_1=0$. Equivalently, one may use $P=0$ while retaining the constraint inherited from its equation. This is the dipole constraint structure of the standard axial formalism \cite{Kase2018}.

\subsection{Direct covariant kinetic calculation}
Define
\begin{equation}
 H=1+4\alpha c,\qquad G=1-4\alpha c.
 \label{eq:HG}
\end{equation}
The angular factor for Eq.~\eqref{eq:oddansatz} is $\int\sin^3\theta\dd\theta\dd\phi=8\pi/3$. Divide out this factor and the positive factor $\mpl^2/2$. In the gauge $P=0$, the terms determining the vector kinetic coefficient are
\begin{equation}
 L_1^{(2)}=
 \frac{r^4H}{2}(\Omega')^2
 -\frac{4\alpha r^2g}{f}\,\dot a_1\Omega'
 +\Omega'\mathcal B[a_1,a_1']+\mathcal C[a_1,a_1'].
 \label{eq:dipoleblock}
\end{equation}
There is no direct $\dot a_1^2$ term. Terms with a single velocity that are irrelevant to the Hessian may be moved by integration by parts. The coefficients in Eq.~\eqref{eq:dipoleblock} follow from direct expansion of Eq.~\eqref{eq:action}. With $P$ retained, the corresponding coefficients of $(\dot P)^2/2$ and $\dot P\dot a_1$ are $r^4H$ and $4\alpha r^2g/f$, respectively. This supplies an independent check before the radial metric gauge is fixed.

For $H\ne0$, solving the rotational constraint in its zero homogeneous sector gives the Schur complement
\begin{equation}
 \Kt=-\frac{8\alpha^2g^2}{Hf^2}.
 \label{eq:Kdip}
\end{equation}
Here $\Kt$ is the coefficient of $\dot a_1^2$ in the angularly reduced action, not the Hessian, which is $2\Kt$. The sign is unaffected by the omitted positive normalization. The derivation does not treat the metric as a fixed external field: the negative term is generated by eliminating its rotational constraint.

The direct radial mixing coefficient is $4\alpha r^2A\,a_1'\Omega'$, and there is no direct $(a_1')^2$ term. Consequently the reduced principal part is
\begin{equation}
 L_{1,\mathrm{pr}}^{(2)}=-\frac{8\alpha^2}{Hf^2}
                 (g\partial_ta_1-fA\partial_ra_1)^2.
 \label{eq:generalprincipal}
\end{equation}
For $H>0$, any real time normal on which the displayed differential direction is nonzero gives a negative kinetic coefficient. Choosing a normal annihilated by that direction makes the vector principal evolution degenerate rather than positive. Thus a regular change of time coordinate cannot supply a positive kinetic form for this principal square.

As a separate convention check, the standard coefficient representation \cite{Kase2018} gives $q_2=(C_1C_5-C_2^2)/C_1$. We divide all reference action coefficients by the common positive factor $p=\mpl^2/2$. The resulting normalized coefficients are
\begin{equation}
 C_1=\frac{H}{2r^2},\qquad C_2=\frac{2\alpha g}{fr^2},\qquad C_5=0,
 \qquad r^2q_2=\Kt.
 \label{eq:Ccheck}
\end{equation}
The Maxwell contribution must be removed when making this comparison, as detailed in Appendix~\ref{app:dictionary}. Keeping a unit Maxwell coefficient would compare a different action.

\subsection{Complete null dipole action and energy sign}
For $c=0$, the lower derivative terms can also be given explicitly. Let
\begin{equation}
 \mathcal Z=u\dot a_1-fu a_1'-f\left(u'-\frac{2u}r\right)a_1.
\end{equation}
On the background relation in Eq.~\eqref{eq:branchrelations}, the direct expansion gives, up to boundary derivatives,
\begin{align}
 L_1^{(2)}&=\frac{r^4}{2}(\Omega')^2
          -\frac{4\alpha r^2}{f}\mathcal Z\Omega'
          +V_1a_1^2,\label{eq:fullnull}\\
 V_1&=4\alpha\left(ru''+2u'-\frac{2u}r\right)
 =-\frac{144\alpha^2u^2(r+2\alpha u)^2}
             {r(r+4\alpha u)^3}.\label{eq:Vdip}
\end{align}
The second form follows by differentiating Eq.~\eqref{eq:firstintegral}. The constraint gives $\Omega'=4\alpha\mathcal Z/(r^2f)$ in the zero angular momentum sector, and hence
\begin{equation}
 L_{1,\mathrm{red}}^{(2)}=-\frac{8\alpha^2}{f^2}\mathcal Z^2+V_1a_1^2.
 \label{eq:nullred}
\end{equation}
This is a complete dipole action on the stated branch, including the potential term.

On an open interval with $\alpha u\ne0$, define $B_1=8\alpha^2u^2/f^2>0$ and
\begin{equation}
 \mathcal D a_1=f a_1'+f\left(\frac{u'}u-\frac2r\right)a_1.
\end{equation}
The momentum is $\pi_1=-2B_1(\dot a_1-\mathcal D a_1)$. The reduced Hamiltonian density is
\begin{equation}
 \mathcal H_1=\pi_1\mathcal D a_1-\frac{\pi_1^2}{4B_1}-V_1a_1^2.
 \label{eq:Hamiltonian}
\end{equation}
Initial data with $a_1=a_1'=0$ and nonzero smooth $\pi_1$ supported in a compact exterior interval have negative quadratic energy. Thus the negative sign is a physical kinetic direction after the constraint is imposed. This statement does not assign an exponential growth rate or a quantum decay rate.

The static coefficient in Eq.~\eqref{eq:Kdip} diverges as $f^{-2}$ near a simple horizon. This factor is not used as evidence for a singular physical rate. Equation~\eqref{eq:EF} provides a regular description of the null background. The energy argument can be made on any compact static interval away from the horizon, where the coordinate transformation is nonsingular.

\section{Tensor characteristics and the incompatibility}\label{sec:tensor}
\subsection{Independent local tensor principal action}
At a regular static point, freeze the background coefficients in a local orthonormal frame. Write $W_{\hat\mu}=(\mathsf A,\mathsf B,0,0)$, where
\begin{equation}
 \mathsf A=\frac A{\sqrt f},\quad \mathsf B=\frac g{\sqrt f},
 \qquad \mathsf A^2-\mathsf B^2=2c.
\end{equation}
A radial helicity two perturbation $h_{23}=h_{32}=\epsilon h(t,x)$ is distinguished from the vector helicities by rotations in the transverse plane. Its principal action therefore supplies a necessary tensor condition. Direct expansion gives
\begin{equation}
 L_T^{(2)}=\frac12\left[(H+4\alpha\mathsf B^2)\dot h^2
 -8\alpha\mathsf A\mathsf B\dot h h_x
 -(G-4\alpha\mathsf B^2)h_x^2\right].\label{eq:TT}
\end{equation}
Only the terms with two derivatives of the perturbation are retained here. The potential and cubic terms do not alter this principal tensor form. Equation~\eqref{eq:TT} follows directly from the curvature expansion; no black hole perturbation coefficient is required to obtain it.

In static coordinates, the radial characteristic polynomial can be written as
\begin{align}
 \mathcal P_T(\omega,k)={}&(fH+4\alpha g^2)\omega^2
 +8\alpha f A g\,\omega k
 -f^2(fG-4\alpha g^2)k^2.\label{eq:PT}
\end{align}
We use the phase convention $e^{i(\omega t-kr)}$; the sign of the mixed term changes if the Fourier convention is reversed, while its discriminant does not. For real nonzero $k$, the discriminant in $\omega$ is
\begin{equation}
 \operatorname{Disc}_\omega\mathcal P_T=4f^4k^2GH.
 \label{eq:tensordisc}
\end{equation}
Real, distinct radial tensor characteristics require $GH>0$. This is necessary, not a sufficient full stability criterion.

For comparison with the spherical formalism, the relevant coefficients are
\begin{align}
 C_8&=-\frac{fG-4\alpha g^2}{2r^4},&
 C_9&=-\frac{4\alpha g}{r^4},&
 C_{10}&=\frac{fH+4\alpha g^2}{2r^4f^2},
\end{align}
which give the independent identity
\begin{equation}
 A^2C_9^2-4f^2C_8C_{10}=\frac{f^2GH}{r^8}.
 \label{eq:sphericaldisc}
\end{equation}
This agrees with the sign of Eq.~\eqref{eq:tensordisc}.

\subsection{A necessary axial obstruction}
\begin{proposition}\label{prop:obstruction}
Consider the action \eqref{eq:action} on a real constant norm background \eqref{eq:branchrelations}. At a regular static point with $f>0$ and $\alpha g\ne0$, there is no nondegenerate axial sector that has both a positive dipole kinetic coefficient and real, distinct radial tensor characteristics.
\end{proposition}
\begin{proof}
Set $\zeta=4\alpha c$. Since $G=1-\zeta$ and $H=1+\zeta$, Eq.~\eqref{eq:tensordisc} requires
\begin{equation}
 GH=1-\zeta^2>0,\qquad -1<\zeta<1.
 \label{eq:strip}
\end{equation}
In this interval $H>0$, and Eq.~\eqref{eq:Kdip} gives $\Kt<0$. Conversely, $\Kt>0$ at a nontrivial point requires $H<0$, hence $\zeta<-1$, for which $GH<0$. The two strict conditions cannot hold together. At $\zeta=\pm1$, the tensor principal form is degenerate, so these values do not satisfy the stated nondegeneracy requirement.
\end{proof}

The obstruction is local. It does not depend on $M$, the size of the horizon, or the detailed location of a circular orbit. For $Q\ne0$, $g=u+cr$ cannot vanish on every exterior open interval. Isolated zeros of $g$ therefore do not produce a globally admissible exterior. The constant $c$ cannot repair the incompatibility by changing the vector while keeping the metric fixed.

The surfaces $H=0$ and $G=0$ require a separate treatment. Section~\ref{sec:boundaries} performs the dipole constraint reduction without dividing by $H$ and examines the tensor principal evolution directly. The trivial $W_\mu=0$ background and the theory at $\alpha=0$ also fall outside Proposition~\ref{prop:obstruction}. The proposition does not address a modified action containing a Maxwell term, additional intrinsic vector interactions, or an extra scalar sector.

\subsection{Sign classification and background checks}\label{sec:benchmarks}
\begin{figure}[htbp]
\centering
\includegraphics[width=\textwidth]{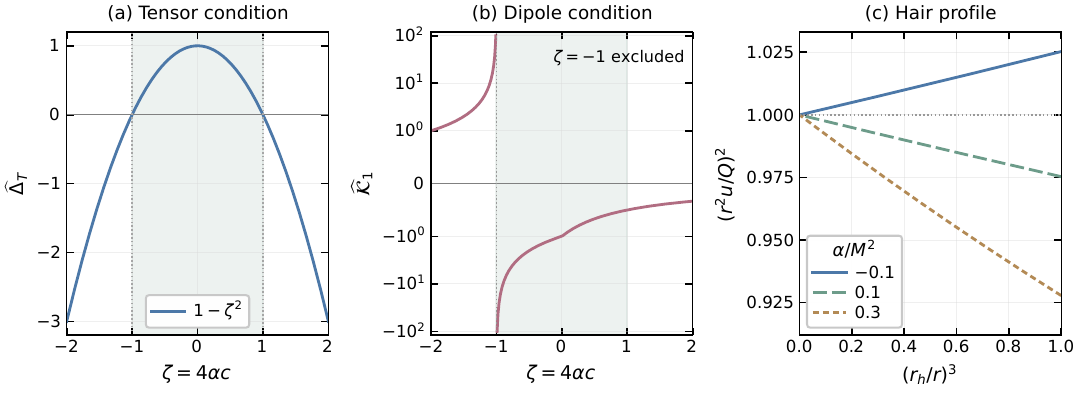}
\caption{Axial consistency diagnostics. (a) The normalized tensor discriminant; the shaded interval has its required positive sign. (b) The dipole kinetic weight on a symmetric logarithmic scale, with the pole at $\zeta=-1$ excluded and its branches drawn separately. The kinetic weight is negative throughout the shaded interval. (c) The shape factor $(r^2u/Q)^2$ for the three null backgrounds in Table~\ref{tab:backgrounds}. The coordinate $(r_h/r)^3$ resolves the leading far deviation and places the horizon at the right edge. The shape factor removes the common leading radial power from the regular weight $8\alpha^2u^2/M^2$ and resolves the coupling dependence; its limit is unity. All curves follow the derived coefficients and exact background.}\label{fig:compatibility}
\end{figure}

Figure~\ref{fig:compatibility} displays the two necessary sign conditions and a regular radial weight on the null branch. The normalized quantities are
\begin{equation}
 \widehat\Delta_T=GH=1-\zeta^2,\qquad
 \widehat{\mathcal K}_1=\frac{f^2\Kt}{8\alpha^2g^2}=-\frac1{1+\zeta}.
\end{equation}
These are dimensionless sign diagnostics. Neither is an observational likelihood or an instability rate. Table~\ref{tab:signs} makes their incompatibility explicit.

\begin{table}[htbp]
\centering\small
\begin{tabular}{llll}
\toprule
Range & $\widehat\Delta_T$ & $\operatorname{sgn}\Kt$ & Necessary condition that fails\\\midrule
$\zeta<-1$ & negative & positive & Real radial tensor characteristics\\
$\zeta=-1$ & zero & see Sec.~\ref{sec:boundaries} & Tensor wave principal form\\
$-1<\zeta<1$ & positive & negative & Positive dipole kinetic energy\\
$\zeta=1$ & zero & negative & Tensor nondegeneracy and dipole sign\\
$\zeta>1$ & negative & negative & Both strict sign requirements\\\bottomrule
\end{tabular}
\caption{Pointwise classification for a real static background with $\alpha g\ne0$. The usual second order dipole coefficient is not assigned at $\zeta=-1$; its regular canonical description is given in Section~\ref{sec:boundaries}.}\label{tab:signs}
\end{table}

The numerical backgrounds in Table~\ref{tab:backgrounds} are auxiliary checks of the analytic identities. For $Q/M=1/2$ and $M=1$, horizon candidates satisfy the exact quartic
\begin{equation}
 r_h^4-2Mr_h^3+\alpha[r_h-2(M-Q)]^2=0.
 \label{eq:quartic}
\end{equation}
The unsquared branch condition must also hold:
\begin{equation}
 1+\frac{2\alpha[r_h-2(M-Q)]}{r_h^3}>0.
\end{equation}
For each rational parameter choice in the table, the largest positive quartic root is isolated in a rational interval of width below $10^{-24}M$. Positivity of the radicand and the unsquared sign is checked throughout that interval using rational lower bounds. The displayed surface gravity $\kappa=f'(r_h)/2$ and the kinetic coefficient are rounded evaluations on that verified horizon branch. They are not independently interval enclosed.

\begin{table}[htbp]
\centering
\begin{tabular}{rrrr}
\toprule
$\alpha/M^2$ & $r_h/M$ & $M\kappa$ & $\Kt(2r_h)/M^2$\\\midrule
$-0.1$ & 2.01257763 & 0.25321413 & $-3.0237298\times10^{-4}$\\
$\phantom{-}0.1$ & 1.98757860 & 0.24696168 & $-3.2287987\times10^{-4}$\\
$\phantom{-}0.3$ & 1.96321555 & 0.24138412 & $-3.0994647\times10^{-3}$\\\bottomrule
\end{tabular}
\caption{Null branch checks with $Q/M=1/2$ and $c=0$. The radicand is real outside the listed simple outer horizon. The negative dipole coefficient is evaluated at $r=2r_h$, away from the horizon coordinate singularity.}\label{tab:backgrounds}
\end{table}

\section{Degenerate boundaries without singular elimination}\label{sec:boundaries}
Appendix~\ref{app:boundaries} retains both rotational variables before gauge fixing and derives the complete action \eqref{eq:completegeneral}. Its velocity Hessian has determinant $-B^2$, with $B=4\alpha r^2g/f$, including at $H=0$. The rotational constraints leave one local canonical pair when $B\ne0$. At $\zeta=-1$, its reduced Hamiltonian \eqref{eq:criticalH} is linear in the surviving momentum and is unbounded below on the compact data described below.

Both tensor principal forms, Eqs.~\eqref{eq:tensorminus} and \eqref{eq:tensorplus}, have rank one. Their first derivative wave reduction has a nontrivial Jordan block and the exact amplification \eqref{eq:amplification}; no frequency independent estimate exists in that wave norm \cite{Sarbach2012}. This is a necessary principal obstruction. The explicit original tensor data and the complete regular Legendre map are given in Appendix~\ref{app:boundaries}. Table~\ref{tab:boundaries} records the physical classification.

\begin{table}[htbp]
\centering\small
\begin{tabular}{p{19mm}p{61mm}p{67mm}}
\toprule
Boundary & Physical dipole at $B\ne0$ & Radial tensor principal system\\\midrule
$\zeta=-1$ & One local canonical pair; $\mathcal H_{1,-}$ is linear in $\pi_1$ and unbounded below & Rank one square, Eq.~\eqref{eq:tensorminus}; defective wave reduction\\[3pt]
$\zeta=1$ & One local canonical pair; $\Kt<0$ & Rank one square, Eq.~\eqref{eq:tensorplus}; defective wave reduction\\
\bottomrule
\end{tabular}
\caption{Boundary classification obtained before division by the vanishing coefficients. The tensor wave reduction assumes a noncharacteristic time normal; otherwise the selected time surface does not define that evolution problem. The count refers to the dipole sector with the global rotational constant fixed.}
\label{tab:boundaries}
\end{table}

\subsection{An explicit compact exterior check}
For an explicit energy example, set $M=1$, $Q/M=1/2$, $\alpha/M^2=1/10$, and $cM^2=-5/2$, so that $\zeta=-1$. The metric is independent of $c$, and its outer horizon is the verified branch in Table~\ref{tab:backgrounds}. Choose $A<0$, which matches $g<0$ at that future horizon. On the interval $3M<r<4M$, the vector is real and the background is regular. Let
\begin{equation}
 a_1(r)=\begin{cases}
  \exp\!\left(1-\dfrac{1}{1-\xi^2}\right),&|\xi|<1,\\
  0,&|\xi|\ge1,
 \end{cases}\qquad
 \xi=\frac{r-7M/2}{M/2}.
 \label{eq:bump}
\end{equation}
This profile and all its derivatives vanish at the endpoints. Set $L_{\rm ref}=M$ in Eq.~\eqref{eq:negativepacket} and choose $\pi_1=-\Lambda_E M^2\mathscr D a_1$, $\Lambda_E>0$, with the transport operator in Eq.~\eqref{eq:Dgeneral}. Define
\begin{equation}
 I_D=\int\dd r\,(\mathscr D a_1)^2,\qquad
 E_0=-\int\dd r\,V_1a_1^2,\qquad
 \Lambda_*=\frac{E_0}{M^2 I_D}.
 \label{eq:packetintegrals}
\end{equation}
Then $E_{1,-}=E_0-\Lambda_E M^2I_D$, with $E_{1,-}<0$ for $\Lambda_E>\Lambda_*$. These data satisfy the rotational constraint because $\dot a_1=\mathscr D a_1$ implies $B\dot a_1+S[a_1]=0$ at $H=0$; the independent momentum $\pi_1$ fixes $\mathcal F_1=\pi_1/B$.

The rounded values are $MI_D=1.09447093945$, $E_0/M=9.97089862687\times10^{-5}$, $\Lambda_*=9.11024520385\times10^{-5}$ and $E_{1,-}(\Lambda_E=1)/M=-1.09437123046$. Multiprecision and independent adaptive quadratures agree at this displayed precision. Restoring the positive factor $(\mpl^2/2)(8\pi/3)$ preserves the sign.

\begin{figure}[htbp]
\centering
\includegraphics[width=\textwidth]{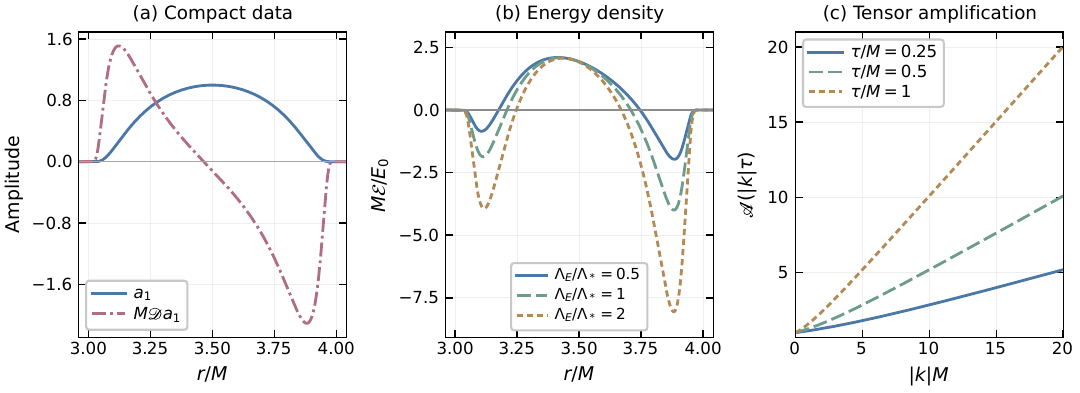}
\caption{Boundary diagnostics at $\alpha/M^2=0.1$, $Q/M=1/2$ and $\zeta=-1$. (a) The smooth amplitude in Eq.~\eqref{eq:bump} and its transport derivative. (b) The local constrained energy density for three momentum families. Its integral is $E_0(1-\Lambda_E/\Lambda_*)$, so a positive local contribution can coexist with a negative total. (c) The exact tensor principal propagator norm $\mathscr A(|k|\tau)$ at three proper times. Its unbounded frequency dependence follows from the repeated characteristic; these curves are not a full black hole evolution.}
\label{fig:boundaries}
\end{figure}

Figure~\ref{fig:boundaries} displays the packet, its constrained energy density $\mathcal E=-\Lambda_E M^2(\mathscr D a_1)^2-V_1a_1^2$, and the tensor amplification. The density integrates to $E_0(1-\Lambda_E/\Lambda_*)$. The amplification is the exact frozen principal result of Eq.~\eqref{eq:amplification}.

\section{A Maxwell deformation of the same vector}\label{sec:maxwell}
\subsection{Modified action and independently varied equations}
Consider the kinetic deformation
\begin{equation}
 I_\eta=\frac{\mpl^2}{2}\int\dd^4x\sqrt{-g}
 \left[R-\alpha\mathcal L_V-\frac\eta4 W_{\mu\nu}W^{\mu\nu}\right],
 \qquad \eta>0,\quad [\eta]=L^2.
 \label{eq:actioneta}
\end{equation}
The field strength here belongs to $W_\mu$ itself. Charged generalizations in Refs.~\cite{Charmousis2025,Charmousis2026} instead introduce an independent electromagnetic field; their solutions do not establish a background for Eq.~\eqref{eq:actioneta}. The remaining vector interactions still break the Maxwell gauge symmetry.

Using the unconstrained spherical ansatz \eqref{eq:ansatz}, the exact radial action is
\begin{equation}
 L_{r,\eta}=L_r+\frac{\eta r^2(a')^2}{2N}.
 \label{eq:radialeta}
\end{equation}
For $E_q=\partial L_r/\partial q-\dd(\partial L_r/\partial q')/\dd r$, its four independently varied equations change by
\begin{align}
 E_{N,\eta}-E_N&=-\frac{\eta r^2(a')^2}{2N^2},&
 E_{f,\eta}-E_f&=0,\nonumber\\
 E_{a,\eta}-E_a&=-\eta\left(\frac{r^2a'}N\right)',&
 E_{b,\eta}-E_b&=0.
 \label{eq:etaadditions}
\end{align}
Equivalently, the covariant vector equation is
\begin{equation}
 \eta\nabla_\nu W^{\nu\mu}-8\alpha\mathcal E_W^\mu=0.
 \label{eq:covarianteta}
\end{equation}
The old background therefore acquires both a temporal vector residual and a lapse residual when it is kept fixed at nonzero $\eta$. The static shift identity in Appendix~\ref{app:variation} remains valid because the added radial term is independent of $b$ and of that shift. Spherical completeness follows from the same covariance identity.

For a variable norm, write $a=NA$, $b=g/f$, and $A^2=g^2+2Xf$. Define $H(X)=1+4\alpha X$ and $G(X)=1-4\alpha X$. Direct combinations of the equations obtained from Eq.~\eqref{eq:radialeta} give, on the radial branch $g\ne0$,
\begin{align}
 E_{f,\eta}-\frac{Xf+g^2}{gf^2}E_{b,\eta}
 &=2r\left[HN'+4\alpha N\left(\frac{2rX}g-1\right)X'\right],
 \label{eq:exactmetriceta}\\
 E_{a,\eta}+\frac{a}{N^2fg}E_{b,\eta}
 &=-\frac{16\alpha ar^2}{Ng}\left(X'+\frac{gN'}{rN}\right)
 -\eta\left(\frac{r^2a'}N\right)'.
 \label{eq:exactvectoreta}
\end{align}
These identities are established before either $X'=0$ or $N'=0$ is imposed.

\subsection{Constant norm rigidity under asymptotic flatness}
\begin{proposition}\label{prop:rigidity}
Let $\alpha\eta\ne0$ and consider a smooth asymptotically flat static exterior of Eq.~\eqref{eq:actioneta}, with $g\ne0$ on a dense subset of the connected exterior extending to infinity. Assume $X=c$, $H(c)\ne0$, and the standard differentiable asymptotic falloff. This radial branch has no nontrivial constant norm vector profile.
\end{proposition}
\begin{proof}
Equation~\eqref{eq:exactmetriceta} forces $N'=0$; normalize $N=1$. The radial vector equation then integrates to
\begin{equation}
 f=1-\frac{2D}r-2ru,\qquad g=u+cr.
 \label{eq:etaradialintegral}
\end{equation}
On these relations the original temporal vector equation vanishes, so Eq.~\eqref{eq:etaadditions} requires
\begin{equation}
 (r^2A')'=0,\qquad A=A_\infty+\frac{P}{r}.
 \label{eq:etacoulomb}
\end{equation}
In particular, $A$ is bounded at infinity. If $f=1-2M/r+O(r^{-2})$, Eq.~\eqref{eq:etaradialintegral} gives $u=(M-D)/r^2+O(r^{-3})$. The exact norm relation is
\begin{equation}
 A^2=(u-cr)^2+2c-\frac{4cD}r.
 \label{eq:etanormrelation}
\end{equation}
Boundedness of $A$ forces $c=0$. Then $A^2=u^2=O(r^{-4})$, and Eq.~\eqref{eq:etacoulomb} implies $A_\infty=P=0$. Hence $A=g=u=0$, contradicting a nontrivial radial profile. The remaining metric is the vacuum solution. Isolated zeros of $g$ do not alter identities that extend by smoothness.
\end{proof}

The proposition concerns this radial branch and the specified action, rather than the broader theories of Ref.~\cite{Heisenberg2017}. At $H=0$, the tensor principal form remains rank one even after the Maxwell term is added, because that term contains no metric derivatives. Thus this excluded algebraic boundary does not satisfy the strict radial tensor conditions either. The limit $\eta\to0$ is singular within the exactly constant norm branch: the Coulomb constraint disappears at $\eta=0$. At $\alpha=0$, outside the proposition, the equations instead admit the Einstein--Maxwell solution $f=1-2M/r+\eta P^2/(4r^2)$, $N=1$, and $a=A_\infty+P/r$; the radial vector component is then gauge freedom.

\subsection{A varying norm response satisfying all radial equations}
To construct a deformation that leaves the rigid branch, expand about the future regular null solution with $Q>0$:
\begin{align}
 X&=\eta\chi+O(\eta^2),& N&=1+\eta n+O(\eta^2),\nonumber\\
 g&=u+\eta j+O(\eta^2),& f&=f_0+\eta h+O(\eta^2),
 \label{eq:responseexpansion}
\end{align}
where $u$ and $f_0$ are Eqs.~\eqref{eq:u} and \eqref{eq:lapse} on the null branch. The dimensions are $[\chi]=L^{-4}$, $[n]=[h]=L^{-2}$, and $[j]=L^{-3}$. Equations~\eqref{eq:exactmetriceta} and \eqref{eq:exactvectoreta}, with vanishing responses at infinity, first give
\begin{equation}
 n=4\alpha\chi,\qquad
 \chi'=-\frac{(r^2u')'}{16\alpha r(r+4\alpha u)}.
 \label{eq:chiODE}
\end{equation}
Set $z=\alpha u/r$. The first integral of the original background implies
\begin{equation}
 z'=-\frac{3z(1+2z)}{r(1+4z)},\qquad
 (r^2u')'=\frac{2u(1-6z-24z^2-8z^3)}{(1+4z)^3}.
 \label{eq:zrelations}
\end{equation}
The solution normalized by $\chi(\infty)=0$ is
\begin{align}
 \chi(r)&=\frac{\Phi(z)}{\alpha^2},\label{eq:closedchi}\\
 \Phi(z)&=\frac{\ln(1+2z)}{48}-\frac{3\ln(1+4z)}{128}
 +\frac{3}{256}\left[1-\frac1{(1+4z)^2}\right].
 \label{eq:Phi}
\end{align}
The logarithms are real on a regular part of the original square root branch, where $z>-1/4$.

The remaining backreaction also has a quadrature representation. Define
\begin{equation}
 J=r(r+4\alpha u)j,\qquad
 S=rh-8\alpha ruj.
 \label{eq:JSdefinitions}
\end{equation}
Both $J$ and $S$ have dimension $L^{-1}$. Independent expansion of the radial vector and lapse equations gives
\begin{align}
 J'={}&\frac{r^2(u')^2}{8}
 +r(3r+4\alpha ru'+8\alpha u)\chi
 -r(r+4\alpha u)\left(\frac{f_0}{u}+4\alpha u\right)\chi',
 \label{eq:Jresponse}\\
 S'={}&-\frac{r^2(u')^2}{4}
 -8\alpha r(ru'+2u)\chi+32\alpha^2ru^2\chi'.
 \label{eq:Sresponse}
\end{align}
Choose $S(\infty)=0$ to keep the ADM mass fixed at this order, and $J(4M)=0$ to specify the homogeneous hair response. Then
\begin{equation}
 j=\frac{J}{r(r+4\alpha u)},\qquad
 h=\frac Sr+8\alpha uj.
 \label{eq:recoverresponse}
\end{equation}
Changing the finite reference condition for $J$ shifts only the homogeneous response. Substitution of Eqs.~\eqref{eq:chiODE}, \eqref{eq:Jresponse}, and \eqref{eq:Sresponse} into all four independently varied field equations cancels every coefficient of order $\eta$.

Regularity is transparent in an ingoing chart. With $\dd v=\dd t+\dd r/(Nf)$ and the future orientation $A\to g$ at a horizon,
\begin{equation}
 \dd s^2=-N^2f\dd v^2+2N\dd v\dd r+r^2\dd\Omega^2,
 \qquad W=NA\dd v-\frac{2X}{g+A}\dd r.
 \label{eq:etaEF}
\end{equation}
The response coefficients remain finite at a regular original horizon with $r+4\alpha u\ne0$ and $u\ne0$. Its first order displacement is $\delta r_h=-\eta h(r_h)/f_0'(r_h)$ for a simple zero.

The expansion is defined on fixed compact exterior intervals. Its formal large radius coefficients are
\begin{equation}
 \chi\sim\frac{Q}{24\alpha r^3},\qquad
 n\sim\frac{Q}{6r^3},\qquad j\sim\frac1{8\alpha r},\qquad
 h\sim\frac Q{r^3}.
 \label{eq:responseasymptotic}
\end{equation}
Consequently $\eta j/u\sim\eta r/(8\alpha Q)$, so an expansion relative to the original vector is not uniform at arbitrarily large radius. A finite $\eta$ global exterior requires a separate far region calculation. The result supplied by Eqs.~\eqref{eq:responseexpansion}--\eqref{eq:recoverresponse} is the complete first order radial response on compact intervals.

\subsection{A common frame for the radial principal forms}
For the modified theory, coordinate signs must be interpreted through the common causal structure \cite{Babichev2018}. Freeze a regular spherical background at a point and use an orthonormal radial frame with $W_{\hat\mu}=(\mathsf A,\mathsf B,0,0)$ and $\mathsf A^2-\mathsf B^2=2X$. A direct local curvature expansion, retaining both rotational shifts before their constraint is imposed, gives
\begin{equation}
 L_{V,\mathrm{pr}}=
 \frac\eta2\left[(\partial_\tau\psi)^2-(\partial_x\psi)^2\right]
 -\frac{8\alpha^2}{H}
       (\mathsf B\,\partial_\tau\psi-\mathsf A\,\partial_x\psi)^2.
 \label{eq:maxwellvectorprincipal}
\end{equation}
Here $\psi$ is a local transverse vector amplitude; a positive local harmonic normalization is understood. The tensor principal action remains Eq.~\eqref{eq:TT}, with $c$ replaced by the pointwise $X$. Terms containing derivatives of the background do not enter these frozen principal forms.

Writing $L_{V,\mathrm{pr}}=\tfrac12K_V\psi_\tau^2+R_V\psi_\tau\psi_x-\tfrac12C_V\psi_x^2$ gives
\begin{align}
 K_V&=\eta-\frac{16\alpha^2\mathsf B^2}{H},&
 C_V&=\eta+\frac{16\alpha^2\mathsf A^2}{H},&
 R_V&=\frac{16\alpha^2\mathsf A\mathsf B}{H},\nonumber\\
 K_VC_V+R_V^2&=\eta\left(\eta+\frac{32\alpha^2X}{H}\right).
 \label{eq:vectorinvariant}
\end{align}
The tensor determinant is $GH$, as before. For $\eta>0$, a common orthonormal radial frame with strictly positive principal Hamiltonians for these two physical sectors exists precisely when
\begin{equation}
 GH>0,\qquad \eta H+32\alpha^2X>0.
 \label{eq:commonframeconditions}
\end{equation}
Indeed, $GH>0$ implies $H,G>0$. For a timelike vector, take its rest frame with $\mathsf B=0$: the tensor weights are $(H,G)$, while the vector weights are $(\eta,\eta+32\alpha^2X/H)$. For a spacelike vector, choose $\mathsf A=0$: the tensor weights become $(G,H)$ and the vector weights are interchanged. For a nonzero null vector, a finite radial boost can make $|\mathsf A|=|\mathsf B|$ sufficiently small that all four weights are positive. Conversely, positivity of both quadratic Hamiltonians requires positive principal determinants and hence Eq.~\eqref{eq:commonframeconditions}.

In the static spherical chart, the vector time coefficient is instead
\begin{equation}
 \mathcal K_{1,\eta}
 =\frac{\eta}{2Nf}-\frac{8\alpha^2g^2}{NHf^2}.
 \label{eq:staticKeta}
\end{equation}
It can remain negative at small $\eta$ even when Eq.~\eqref{eq:commonframeconditions} holds in a common frame. This does not contradict the original obstruction: at $\eta=0$ and $H>0$, Eq.~\eqref{eq:maxwellvectorprincipal} is a negative principal square on every noncharacteristic time normal. With $\eta>0$, its causal structure changes. Equation~\eqref{eq:commonframeconditions} classifies the frozen radial tensor and transverse vector forms. Section~\ref{sec:angular} supplies the separate even angular test.

\subsection{Numerical backreaction and full equation residuals}
We evaluate the response on the three original backgrounds in Table~\ref{tab:backgrounds}, with the integration conditions stated after Eq.~\eqref{eq:Sresponse}. The quantity
\begin{equation}
 \mathcal C_0=1+32\alpha^2\chi
 \label{eq:leadingmargin}
\end{equation}
is the leading value of $(\eta H+32\alpha^2X)/\eta$ along the response. At $r=2r_h$ its values range from $0.99896827$ to $1.00320790$ for the three backgrounds, and are positive.

Two independent integrations determine the responses: adaptive integration of Eqs.~\eqref{eq:Jresponse} and \eqref{eq:Sresponse}, and multiprecision quadrature of their sources. The differential integration uses relative and absolute tolerances $2\times10^{-13}$ and $2\times10^{-15}$ for $MJ$ and $MS$; the independent quadrature uses 70 decimal digits. Their absolute differences are below $2.5\times10^{-14}$ at the comparison points. To test the full equations, reconstruct
\begin{equation}
 \begin{gathered}
 N_\eta=1+4\alpha\eta\chi,\quad f_\eta=f_0+\eta h,\quad
 g_\eta=u+\eta j,\\
 X_\eta=\eta\chi,\quad
 a_\eta=N_\eta\sqrt{g_\eta^2+2X_\eta f_\eta},\quad b_\eta=g_\eta/f_\eta.
 \end{gathered}
 \label{eq:reconstructedeta}
\end{equation}
The exact square root here enforces the norm of the reconstructed fields; it does not make their field equations exact beyond first order. Evaluate all four full equations from Eq.~\eqref{eq:radialeta}, including their Maxwell terms, at $r/r_h=1.2,2,4$. Define the dimensionless residual norm
\begin{equation}
 \mathscr R_{\max}=\max_{r/r_h,\,q}
  \left\{|E_{N,\eta}|,|E_{f,\eta}|,
               |E_{a,\eta}|/M,|E_{b,\eta}|/M\right\}.
 \label{eq:residualnorm}
\end{equation}
For successive halvings of $\eta/M^2$ from $4\times10^{-4}$ to $2.5\times10^{-5}$, the observed residual orders lie between $1.9995$ and $2.0006$, consistent with the independent cancellation of every linear coefficient.

\begin{figure}[htbp]
\centering
\includegraphics[width=\textwidth]{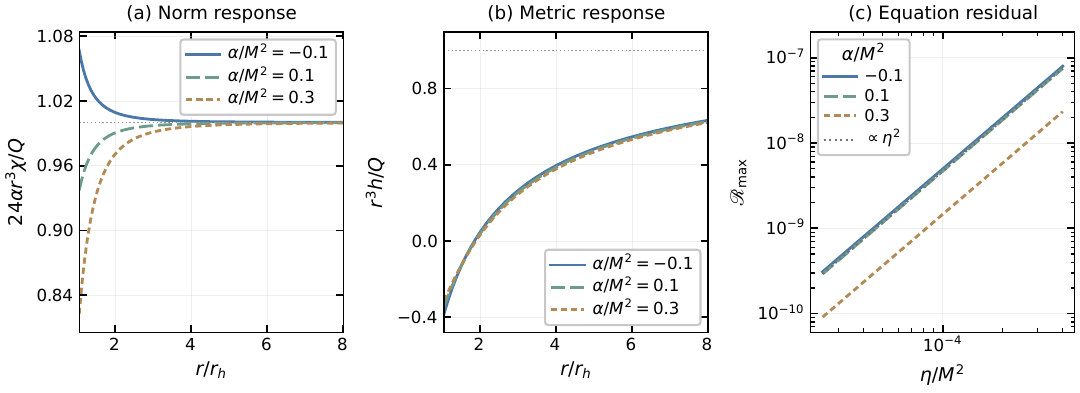}
\caption{First order Maxwell response from the independently varied radial equations. (a) The scaled norm response $24\alpha r^3\chi/Q$. (b) The scaled metric response $r^3h/Q$. Both tend to unity by Eq.~\eqref{eq:responseasymptotic}, and use $Q/M=1/2$, $S(\infty)=0$, $J(4M)=0$. The scaling separates the background dependence from the common radial powers. (c) Maximum residual of all four full radial equations on Eq.~\eqref{eq:reconstructedeta}, at the stated test radii. The dotted line is the quadratic scaling expected for a first order background.}
\label{fig:maxwell}
\end{figure}

Figure~\ref{fig:maxwell} shows the scaled norm and metric responses, whose leading far limits are unity, together with the full equation convergence test. At $\eta/M^2=10^{-4}$, the sampled reconstructed profiles over $1.05\le r/r_h\le8$ have positive $H$, $G$, and $(\eta H+32\alpha^2X)/\eta$. The largest sampled value of $|\eta n|$, $|\eta h/f_0|$, $|\eta j/u|$, and $|\eta\chi f_0/u^2|$ is below $2.51\times10^{-3}$. These checks support the first order response and radial principal compatibility at the sampled points; they do not bound the solution error throughout the interval.

\section{Far boundary conditions and the singular outer problem}\label{sec:far}
\subsection{Absence of a nonzero Coulomb temporal coefficient}
The loss of uniformity in Eq.~\eqref{eq:responseasymptotic} requires an independent asymptotic calculation. We first ask whether the addition of a Maxwell term permits a Coulomb potential that vanishes at infinity. Assume a standard asymptotically flat end,
\begin{equation}
 N=1+O(r^{-2}),\qquad f=1-\frac{2M}{r}+O(r^{-2}),
 \label{eq:farAF}
\end{equation}
with the corresponding first and second derivative falloffs. Impose
\begin{equation}
 a=\frac{P}{r}+o(r^{-1}),\qquad g=\frac{B}{r}+o(r^{-1}),
 \label{eq:farCoulomb}
\end{equation}
again with differentiated remainders through the order required by the field equations. The assumption includes a faster radial falloff through $B=0$; it does not prescribe every possible vector asymptotic behavior.

The exact flat metric vector equations provide a transparent way to extract the leading coefficients. Setting $N=f=1$ only after varying the full radial action gives
\begin{align}
 \mathscr E_r&=2aa'+\frac{4g^2}{r}+3(g^2-a^2)g=0,\label{eq:flatfarR}\\
 \mathscr E_t&=\eta\left(a''+\frac{2a'}r\right)
 -8\alpha a\left[2g'+\frac{4g}r+3(g^2-a^2)\right]=0.
 \label{eq:flatfarT}
\end{align}
These follow independently from $E_b/r^2=-8\alpha\mathscr E_r$ and $E_a/r^2=-\mathscr E_t$ in the flat metric. Under Eqs.~\eqref{eq:farAF} and \eqref{eq:farCoulomb}, the curvature and metric corrections do not contribute to the coefficients of $r^{-3}$ in these vector equations. The quartic interaction does contribute and must be retained.

For $\alpha\ne0$ and $P\ne0$, the two leading coefficients require
\begin{align}
 R_C&=4B^2-2P^2+3B(B^2-P^2)=0,\nonumber\\
 T_C&=2B+3(B^2-P^2)=0.\label{eq:Coulombcoefficients}
\end{align}
But $R_C-BT_C=2(B^2-P^2)$ implies $B^2=P^2$, whereupon $T_C=2B$ gives $B=P=0$. This contradicts $P\ne0$. Thus no solution in this asymptotic class has a nonzero Coulomb temporal coefficient. The conclusion is an asymptotic restriction on this action, rather than a general no hair theorem. In particular, when $P=0$, the leading radial equation also permits $B=-4/3$; that branch is not classified here. The Einstein--Maxwell limit $\alpha=0$ is separately covered by the exact check in Section~\ref{sec:maxwell} and is not subject to the division by $\alpha$ used in Eq.~\eqref{eq:Coulombcoefficients}.

\subsection{Outer scaling and overlap with the compact response}
For the connection to the $Q>0$ future regular null seed, take $\alpha>0$ and define
\begin{equation}
 \lambda=\frac{\eta}{\alpha}>0,\qquad
 \rho=\frac{\lambda r}{Q},\qquad
 a=\frac{\lambda^2}{Q}\mathcal A(\rho),\qquad
 g=\frac{\lambda^2}{Q}\mathcal B(\rho).
 \label{eq:outerscaling}
\end{equation}
All four quantities $\lambda,\rho,\mathcal A,\mathcal B$ are dimensionless. The crossover scale is $r\sim Q/\lambda=\alpha Q/\eta$. In the weak deformation limit with fixed $\rho$, the Maxwell and cubic terms balance. The quartic vector terms are smaller by one power of $\lambda$ at fixed $\rho$, and weak geometry corrections vanish in this limit. Direct substitution into the full vector Euler equations verifies the same limit for $f=1-2M/r+F_2/r^2$ and $N=1+N_2/r^2$ with bounded coefficients. This reduction supplies the leading outer vector equations; it does not impose the full metric equations.

The resulting system is
\begin{equation}
 \mathcal A\mathcal A_{,\rho}=-\frac{2\mathcal B^2}{\rho},\qquad
 \mathcal A_{,\rho\rho}+\frac{2\mathcal A_{,\rho}}{\rho}
 =16\mathcal A\left(\mathcal B_{,\rho}+\frac{2\mathcal B}{\rho}\right).
 \label{eq:outervector}
\end{equation}
For $t=\ln\rho$, $q=\rho\mathcal A$, $k=\rho\mathcal B$, and $y=k/q$, it becomes an autonomous first order system,
\begin{align}
 \dot q&=q(1-2y^2),\nonumber\\
 \dot y&=-\frac{y\left[y(1-2y^2)+16q(1-y^2)\right]}{2(y+4q)}.
 \label{eq:outerphase}
\end{align}
Dots in this section denote derivatives with respect to $t$, not physical time. Before eliminating $k$, the temporal equation is $\ddot q-\dot q=16q(\dot k+k)$.

At fixed $M/Q$ and $\alpha/Q^2$, a shared weak deformation overlap has $\lambda\ll\rho\ll1$: the geometry is weak there while $\eta j/u\sim\rho/8$ is small. The inner overlap expansion of Eq.~\eqref{eq:outervector} is
\begin{align}
 q={}&\frac1\rho+\frac16-\frac{11\rho}{576}
       +\frac{5\rho^2}{864}
       +\rho^3\left(\frac{5\ln\rho}{27648}+C_{\rm in}\right)+\cdots,
       \label{eq:innerouterq}\\
 k={}&\frac1\rho+\frac18-\frac\rho{96}+\frac{\rho^2}{512}
       -\frac{19\rho^3}{55296}+\cdots.
       \label{eq:innerouterk}
\end{align}
Substitution into both original equations in Eq.~\eqref{eq:outervector} independently fixes the displayed logarithm. The undetermined coefficient $C_{\rm in}$ represents additional boundary data in the leading outer problem. In particular,
\begin{equation}
 a=\frac Q{r^2}+\frac{\eta}{6\alpha r}+\cdots,\qquad
 g=\frac Q{r^2}+\frac{\eta}{8\alpha r}+\cdots
 \label{eq:outerfirstmatch}
\end{equation}
in the overlap. The temporal coefficient agrees with $j+\chi f_0/u+nu$ from Eq.~\eqref{eq:responseexpansion}; using $j$ alone would give the wrong factor. The coefficient $C_{\rm in}$ enters $\mathcal A=q/\rho$ at order $\rho^2$, and therefore cannot be fixed by the first order compact response. Equations~\eqref{eq:innerouterq} and \eqref{eq:innerouterk} are overlap series of the leading outer vector system. They are not higher order solutions of the full coupled gravitational problem.

\subsection{Logarithmic branch and vanishing asymptotic flux}
The positive slow branch at $q\to0$ has
\begin{equation}
 y=\frac1{\sqrt2}+4q-88\sqrt2q^2+O(q^3),\qquad
 \dot q=-8\sqrt2q^2+320q^3+O(q^4)
 \label{eq:slowouter}
\end{equation}
for the leading outer equations. The same leading logarithm also follows with the quartic interaction retained at fixed $\lambda$. Up to terms suppressed by the asymptotically flat metric corrections, the complete far vector balance is
\begin{align}
 q(\dot q-q)+2k^2+\frac{3\lambda}{2}k(k^2-q^2)&=0,\nonumber\\
 \ddot q-\dot q-16q(\dot k+k)-24\lambda q(k^2-q^2)&=0.
 \label{eq:finitefarbalance}
\end{align}
Its formal slow branch satisfies
\begin{align}
 y&=\frac1{\sqrt2}+\left(4+\frac{3\lambda}{16}\right)q+O(q^2),\nonumber\\
 \dot q&=-8\sqrt2q^2+(320+9\lambda)q^3+O(q^4).
 \label{eq:finitefarbeta}
\end{align}
Thus a slowly varying positive asymptotic branch has
\begin{equation}
 \frac1q=8\sqrt2\,t-\frac{320+9\lambda}{8\sqrt2}\ln t+O(1),
 \qquad t\to\infty.
 \label{eq:logrunning}
\end{equation}
The leading fields can be written using a fixed positive reference length $r_*$ as
\begin{align}
 a&\sim\frac{\eta}{8\sqrt2\alpha\,r\ln(r/r_*)},&
 g&\sim\frac{a}{\sqrt2},&
 X&\sim\frac{\eta^2}{512\alpha^2r^2\ln^2(r/r_*)}.
 \label{eq:logfields}
\end{align}
The reference scale in this leading asymptotic form is fixed by boundary matching beyond leading order. These are formal asymptotic consequences of the far vector equations under Eq.~\eqref{eq:farAF}; they do not by themselves prove a horizon to infinity solution.

The electric flux of the same vector,
\begin{equation}
 \mathcal Q_E(r)=-\frac{r^2a'}{N}
   \sim\lambda(q-\dot q)
   \sim\frac{\eta}{8\sqrt2\alpha\ln(r/r_*)},
 \label{eq:runningflux}
\end{equation}
tends to zero. It is not a conserved Maxwell charge because the remaining vector interactions break the gauge symmetry. The logarithm therefore evades the nonzero Coulomb coefficient restriction without supplying a new asymptotic electric charge. The norm is positive and vanishes at infinity on this branch. Consequently the two radial principal conditions in Eq.~\eqref{eq:commonframeconditions} are met sufficiently far out for fixed $\eta>0$. The separate longitudinal calculation in Section~\ref{sec:angular} determines the angular sign; the known cubic theory analysis provides a useful comparison \cite{DeFelice2024}.

\subsection{Computed connection of the leading outer equations}
We calculate the positive connection of Eq.~\eqref{eq:outerphase} in the phase variable $s=\ln q$. The slow branch supplies $y$ at $q_{\min}=10^{-4}$ through a formal slow expansion of order $q^8$ with exact coefficients. Integrating toward $q_{\max}=10^4$ proceeds backwards in physical log radius and avoids forward amplification away from the slow branch. The normalization inherited from the seed is $\rho q\to1$ in the inner overlap. At the large $q$ endpoint, we impose its independently derived expansion,
\begin{equation}
 \ln(\rho q)=\frac1{6q}-\frac1{192q^2}+\frac5{5184q^3}
 +O\!\left(\frac{\ln q}{q^4}\right).
 \label{eq:radiusnormalization}
\end{equation}
The differential equations used for the phase integration are $\dd y/\dd s=\dot y/(1-2y^2)$ and $\dd\ln(\rho q)/\dd s=1+(1-2y^2)^{-1}$.

Table~\ref{tab:farconnection} gives representative values of the normalized connection. The primary integration uses relative tolerance $3\times10^{-14}$, absolute tolerances $10^{-15}$ and $5\times10^{-14}$ for $y$ and the integrated radius variable, and maximum step $0.04$ in $s$. An independent implicit integration agrees at the tabulated points within $2.5\times10^{-11}$ in relative $q$ and $4.7\times10^{-13}$ in $y$. Repeating the calculation with $q_{\min}=2\times10^{-4}$ and $q_{\max}=2\times10^4$, or with the initial slow series truncated at order $q^6$, changes $q$ there by less than $4.7\times10^{-13}$ in relative value. These are numerical comparisons, rather than rigorous error enclosures.

\begin{table}[htbp]
\centering\small
\begin{tabular}{rrrr}
\toprule
$t=\ln\rho$ & $q$ & $y=k/q$ & $-\dot q/(8\sqrt2q^2)$\\\midrule
$0$  & $1.15171615$   & $0.96917564$ & $0.06742829$\\
$5$  & $0.0614245113$ & $0.81726051$ & $0.48325031$\\
$10$ & $0.0196721765$ & $0.76053545$ & $0.70463980$\\
$20$ & $0.00711153913$& $0.73093677$ & $0.85183857$\\
$40$ & $0.00290621161$& $0.71782364$ & $0.92887890$\\
$80$ & $0.00129063189$& $0.71207596$ & $0.96593025$\\
\bottomrule
\end{tabular}
\caption{Computed connection of the leading outer vector equations \eqref{eq:outerphase}, normalized by Eq.~\eqref{eq:radiusnormalization}. The last two columns tend to $1/\sqrt2$ and $1$ on the slow branch. These values describe the dimensionless outer problem, rather than a full finite $\eta$ black hole background.}
\label{tab:farconnection}
\end{table}

An independent integral diagnostic follows by defining $\mathcal F_E=2k^2/q=q-\dot q$:
\begin{equation}
 \frac{\dd}{\dd t}\left(\mathcal F_E+16qk\right)
 =-16k\mathcal F_E.
 \label{eq:outerfluxidentity}
\end{equation}
Adaptive quadrature over four intervals between $t=-5$ and $80$ verifies its normalized integral defect below $10^{-13}$. Five point differences in the original radial and second order temporal equations give normalized residuals below $3.5\times10^{-11}$ and $1.6\times10^{-9}$ at $t=0,5,10,20$. Direct evaluation of the omitted flat quartic terms also confirms their linear scaling with $\lambda$ at fixed $\rho$; it does not test the metric equations.

\begin{figure}[htbp]
\centering
\includegraphics[width=\textwidth]{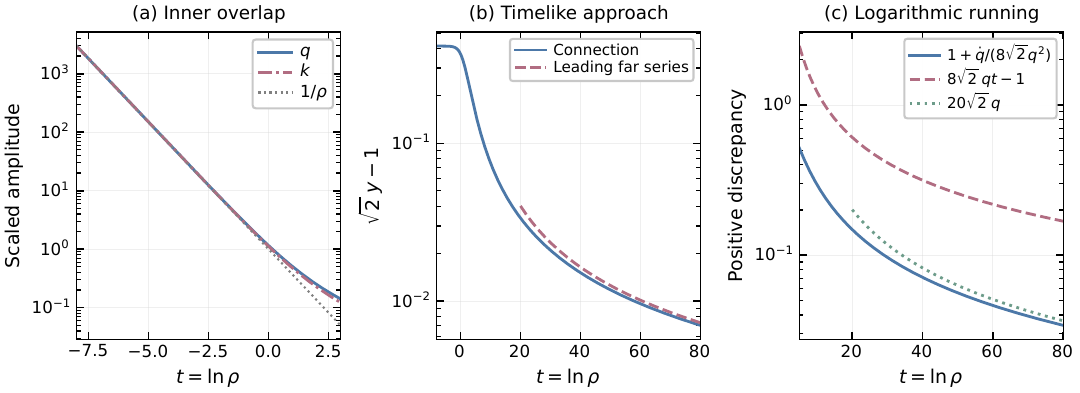}
\caption{Leading outer vector connection normalized by Eq.~\eqref{eq:radiusnormalization}. (a) The temporal and radial amplitudes on the inner overlap range, with $1/\rho$ displayed over that same range. (b) The positive deviation $\sqrt2\,y-1$ tends to zero as the far vector becomes timelike. The leading far series $4\sqrt2\,q$ is shown for $t\ge20$. (c) Positive deviations from the two logarithmic limits: $1+\dot q/(8\sqrt2q^2)$ and $8\sqrt2qt-1$. The dotted curve $20\sqrt2q$, also shown for $t\ge20$, is the first slow series correction to the former. The logarithmic ordinates resolve the full displayed tails. These are leading outer vector results for weak deformation.}
\label{fig:farconnection}
\end{figure}

Figure~\ref{fig:farconnection} displays the vector connection and its approach to the logarithmic balance. The ratio $8\sqrt2qt$ converges slowly because of the logarithm of $t$ in Eq.~\eqref{eq:logrunning}. The boundary calculation identifies a compatible far vector behavior and an additional matching requirement. Joining this far behavior to a regular horizon requires the full finite $\eta$ gravitational calculation given in Section~\ref{sec:finite}.

\section{Finite deformation exteriors and regular horizons}\label{sec:finite}
\subsection{Ingoing variables and the algebraic constraint}
To impose the metric equations at finite $\eta$, use the regular chart
\begin{equation}
 \dd s^2=-N^2f\dd v^2+2N\dd v\dd r+r^2\dd\Omega^2,
 \qquad W=a\dd v+w\dd r,
 \label{eq:finiteingoing}
\end{equation}
where $A=a/N$, $g=A+fw$, and $X=-Aw-fw^2/2$. The original diagonal variable is $b=a/(Nf)+w$. Substituting this relation into the independently varied radial action gives a regular Lagrangian of the form \eqref{eq:radialeta}, with
\begin{equation}
 Y=2Aw+fw^2,\qquad
 \Theta=g'+\left(\frac{N'}N+\frac2r\right)g,\qquad
 \mathcal I=Y\frac{rf'+f-1}{r^2}+\frac{2g^2N'}{rN}.
 \label{eq:regularradialterms}
\end{equation}
No inverse power of $f$ remains. A separate variation of this regular action agrees with all four original Euler equations through
\begin{align}
 \widetilde E_N&=E_N-\frac{a}{N^2f}E_b,&
 \widetilde E_f&=E_f-\frac{a}{Nf^2}E_b,\nonumber\\
 \widetilde E_a&=E_a+\frac{E_b}{Nf},&
 \widetilde E_w&=E_b.
 \label{eq:ingoingchain}
\end{align}
The displayed combinations extend regularly to the horizon. The shift and angular identities in Appendix~\ref{app:variation} still account for spherical completeness.

The coefficient matrix of $N'/N$, $f'$, and $w'$ in $(\widetilde E_N,\widetilde E_f,\widetilde E_w)$ has a left null vector. Its compatibility equation is algebraic. To display it compactly and retain the ADM mass at very large radius, introduce
\begin{align}
 t&=\ln(r/M),& m&=\frac r2(1-f),&\sigma&=\ln N,\nonumber\\
 p&=rg,&z&=r^2X,&\mathfrak a&=rA=\sqrt{p^2+2fz},\nonumber\\
 \mathfrak e&=\frac{r^2a'}N,&\beta&=\frac\alpha{r^2},&\lambda&=\frac\eta\alpha.
 \label{eq:finitevariables}
\end{align}
All variables in Eq.~\eqref{eq:finitevariables} except $m$ are dimensionless. Dots below denote derivatives with respect to this radial $t$. Define
\begin{equation}
 H=1+4\beta z,\qquad \mathcal D=1+4\beta(p-z),\qquad
 \mathcal T=p^2+3pz-6z^2-2(1-f)z.
 \label{eq:finiteHD}
\end{equation}
The exact compatibility equation is
\begin{align}
 \mathcal C={}&\frac{\lambda\beta}{8}(p-2z)\mathfrak e^2
 -\mathfrak a\mathcal D\mathfrak e
 -2\beta(p-z)\mathcal T\nonumber\\
 &-p(2p-3z)+(1-f)z=0.
 \label{eq:finiteconstraint}
\end{align}
It is derived before imposing a norm or a relation between the two metric functions. Together with the metric and radial vector equations, it implies the lapse equation on the branch $p-2z\ne0$.

Write $\mathcal C=c_2\mathfrak e^2+c_1\mathfrak e+c_0$ and $\Delta=c_1^2-4c_2c_0$. The root connected to the decreasing null seed is evaluated as
\begin{equation}
 \mathfrak e=\frac{2c_0}{\mathfrak a\mathcal D+\sqrt\Delta},\qquad
 \mathcal C_{,\mathfrak e}=-\sqrt\Delta.
 \label{eq:electricroot}
\end{equation}
This rationalized expression avoids subtracting nearly equal terms and has a finite limit when $c_2\to0$. We use the real branch with $\mathfrak a,p,H,\mathcal D,p-2z>0$ and $\Delta>0$.

\subsection{Exact evolution and a simple horizon}
Set
\begin{equation}
 V=\frac{4\beta(1-2z/p)}H,\qquad
 S=\frac{\beta}{\mathcal D}
       \left[\frac{\lambda\mathfrak e^2}{4}+12(p-z)^2\right].
 \label{eq:finiteVS}
\end{equation}
The full equations and the derivative of the norm relation reduce to
\begin{align}
 \dot\sigma&=V(\dot z-2z),\nonumber\\
 \dot f+2f\dot\sigma&=1-f-S,\nonumber\\
 p\dot p+f\dot z+p^2\dot\sigma
 &=\mathfrak a\mathfrak e+p^2+(3f-1)z+zS,\nonumber\\
 \frac{\mathcal D}{H}(\dot z-2z)
    +\frac{\lambda p}{16\mathfrak a}\dot{\mathfrak e}&=0,
 \qquad \dot{\mathcal C}=0.
 \label{eq:finiteflow}
\end{align}
These equations determine the derivative of the algebraic root along the solution, rather than treating $\mathfrak e$ as unconstrained data.

For an explicit first order representation, write $\dot f=F_0+F_1\dot z$, $\dot p=P_0+P_1\dot z$, and $\dot{\mathfrak e}=E_0+E_1\dot z$, where
\begin{align}
 F_0&=1-f-S+4fzV,&F_1&=-2fV,\nonumber\\
 P_0&=\frac{\mathfrak a\mathfrak e+p^2+(3f-1)z+zS+2zp^2V}{p},&
 P_1&=-\frac{f+p^2V}{p},\nonumber\\
 E_1&=-\frac{16\mathfrak a\mathcal D}{\lambda pH},& E_0&=-2zE_1.
 \label{eq:finiteaffine}
\end{align}
Partial derivatives of $\mathcal C$ include the dependence $\mathfrak a=\sqrt{p^2+2fz}$. Since $\dot\beta=-2\beta$, define
\begin{align}
 \mathcal J&=\mathcal C_{,f}F_1+\mathcal C_{,p}P_1
                 +\mathcal C_{,z}+\mathcal C_{,\mathfrak e}E_1,\nonumber\\
 \mathcal R&=\mathcal C_{,f}F_0+\mathcal C_{,p}P_0
                 +\mathcal C_{,\mathfrak e}E_0-2\beta\mathcal C_{,\beta},
 \qquad \dot z=-\frac{\mathcal R}{\mathcal J}.
 \label{eq:finiteJacobian}
\end{align}
Thus $m,p,z,\sigma$ obey an explicit system wherever the stated branch conditions and $\mathcal J\ne0$ hold. The denominator $\mathcal J$ tests this static reduction; it is not a substitute for a perturbation characteristic or energy condition.

We evolve the mass directly through
\begin{equation}
 \dot m=\frac\alpha r\left[
 \frac{4f(1-2z/p)}H(\dot z-2z)
 +\frac{\lambda\mathfrak e^2/8+6(p-z)^2}{\mathcal D}\right].
 \label{eq:finitemass}
\end{equation}
This retains $M$ even where $1-2m/r$ rounds to unity. Independent substitutions verify the regular action and all reduction identities. The $\beta\to0$ limit reproduces Eq.~\eqref{eq:finitefarbalance}; at $\eta=0$, treated separately from division by $\lambda$, the constraint recovers the original null seed relation for $u'$.

The metric tail also follows directly from independent variation of Eq.~\eqref{eq:radialeta}. Since $\dot q=O(q^2)$, set $a=\lambda q/r$, $g=\lambda q/(\sqrt2r)$ at the first nonzero slow order. Substituting $N=1+n_2\alpha\lambda^2q^2/r^2$ and $f=1-2M/r+d_2\alpha\lambda^2q^2/r^2$ gives $n_2=1$ and $d_2=2+\lambda/4$. Thus, at fixed positive $\lambda$,
\begin{align}
 N-1&=\frac{\alpha\lambda^2q^2}{r^2}\,[1+O(q)+O(M/r)],\nonumber\\
 m-M&=-\frac{\alpha\lambda^2}{r}\left(1+\frac\lambda8\right)q^2
                    [1+O(q)+O(M/r)].
 \label{eq:gravitationaltails}
\end{align}
The mass integral converges and both metric functions have the assumed asymptotically flat end. Normalized differential tails on the full backgrounds approach these coefficients; this comparison avoids subtracting the unresolved tail from $M$.

At a zero of $f$ with $p>0$, the ingoing vector has $A=g$ and $w=-2X/(g+A)$ finite. If the reduction remains regular and $S_h<1$, the zero is simple:
\begin{equation}
 \dot f_h=1-S_h>0,\qquad
 \kappa=\frac{N_h(1-S_h)}{2r_h}.
 \label{eq:finitehorizon}
\end{equation}
The smooth ingoing flow then extends locally across the horizon. For example, with $\nu=N'/N$, its Kretschmann scalar is
\begin{equation}
 K_h=4\left[\left(\frac{f_h''}{2}+\frac32\nu_h f_h'\right)^2
                 +\frac{(f_h')^2}{r_h^2}+\frac1{r_h^4}\right],
 \label{eq:finitecurvature}
\end{equation}
which follows from an independent contraction of the four dimensional Riemann tensor. It is evaluated without a static coordinate pole and reduces to $48M^2/r_h^6$ in the Schwarzschild limit.

\subsection{Asymptotic normalization and numerical exteriors}
The far boundary must be imposed consistently when its starting point is changed. With $\mathcal U=\sqrt2q$ and $\mathcal Y=\sqrt2y$, Eq.~\eqref{eq:finitefarbalance} gives a formal asymptotic slow expansion with exact recursively determined coefficients. Its radius relation is
\begin{equation}
 T_\lambda(\mathcal U)=\frac1{8\mathcal U}
 -\left(\frac52+\frac{9\lambda}{128}\right)\ln \mathcal U
 +\sum_{n=1}^{6}t_n(\lambda)\mathcal U^n+O(\mathcal U^7).
 \label{eq:finitefarseries}
\end{equation}
The coefficients follow from the slow series for $\mathcal Y$ through order $\mathcal U^8$ and $\dd T_\lambda/\dd \mathcal U=1/\dot{\mathcal U}$.  This relation includes the quartic correction to the logarithm in Eq.~\eqref{eq:logrunning}.

Choose a far length convention by
\begin{equation}
 \ln(r/M)=\ln\!\left(\frac{Q_*}{M\lambda}\right)+\ell_0+T_\lambda(\mathcal U),
 \qquad \ell_0\simeq-4.33826505.
 \label{eq:farlengthconvention}
\end{equation}
The leading outer system fixes $\ell_0$ by $\rho q\to1$ in its inner overlap, as in Section~\ref{sec:far}. The parameter $Q_*$ labels this length convention; it is not a new conserved electric charge. We fix $Q_*/M=1/2$ and $\alpha/M^2=0.1$. For this convention, the numerical norm and lapse approach the original null seed response as $\lambda$ decreases, as checked below. Fixing $Q_*$ differs from imposing $J(4M)=0$ at first order.

At $q_f=2\times10^{-4}$, set $m_f=M$ and $\sigma_f=0$, with
\begin{equation}
 p_f=\lambda q_f y_f,\qquad
 z_f=\frac{\lambda^2q_f^2(1-y_f^2)}{2f_f},\qquad
 f_f=1-\frac{2M}{r_f}.
 \label{eq:finiteinitialdata}
\end{equation}
The omitted asymptotic mass and lapse tails vanish as inverse powers of $r_f$; the chosen far points have $\ln(r_f/M)>459$. We integrate inward using $m/M,p,z,\sigma$, eliminating the electric root at every step. The primary relative tolerance is $2\times10^{-12}$, with absolute tolerances $(2\times10^{-14},2\times10^{-18},2\times10^{-22},2\times10^{-16})$ and maximum step $0.15$ in $\ln(r/M)$.

Each of the four couplings in Table~\ref{tab:finitehorizons} reaches a simple future horizon while remaining on the declared real branch. The minimum sampled $\mathcal J$ exceeds $4.0045$, and all sampled branch and normalized radial principal margins are positive. The rescaled norm $z$ remains positive even where the unscaled $X$ would underflow in the very far region. Thus these calculations construct finite deformation exteriors matched to the logarithmic far branch, with both metric functions determined by the full equations.

\begin{table}[htbp]
\centering\small
\begin{tabular}{rrrrrr}
\toprule
$\lambda$ & $r_h/M$ & $10^6(N_h-1)$ & $10^6M^2X_h$ & $M\kappa$ & $M^4K_h$\\\midrule
$0.1$   & $1.983595135$ & $108.129412$ & $272.396229$ & $0.2460642453$ & $0.728312182$\\
$0.03$  & $1.986121380$ & $30.767171$  & $77.101026$  & $0.2466262088$ & $0.731171676$\\
$0.01$  & $1.987000900$ & $10.021135$  & $25.072954$  & $0.2468271497$ & $0.732220025$\\
$0.003$ & $1.987373061$ & $2.973548$   & $7.435669$   & $0.2469134271$ & $0.732674909$\\
\bottomrule
\end{tabular}
\caption{Numerically constructed full finite deformation backgrounds at $\alpha/M^2=0.1$ and $Q_*/M=1/2$, with the far length convention \eqref{eq:farlengthconvention} and asymptotic normalization $N\to1$. The curvature and surface gravity use Eqs.~\eqref{eq:finitehorizon} and \eqref{eq:finitecurvature}. These are finite $\eta=\alpha\lambda$ solutions, rather than first order reconstructed fields.}
\label{tab:finitehorizons}
\end{table}

\subsection{Independent equation and convergence checks}
An independent implicit integration agrees with the primary integration at $r/M=2.1,4,10,100$ within $5.6\times10^{-11}$ in relative $m,p,z$. The largest relative horizon change among this comparison, tighter tolerances, the endpoint $q_f=3\times10^{-4}$, and a slow series truncated at order $\mathcal U^6$ is below $1.1\times10^{-11}$. The far length convention is held fixed through Eq.~\eqref{eq:finitefarseries}; changing $q_f$ without that radius adjustment would instead change the physical boundary data.

We also substitute the profiles into all four independently varied regular Euler equations at eight radii, including the horizon. With derivatives from the differential flow, the algebraic residuals are at rounding precision, consistently with the exact reduction. A separate check obtains every first and second derivative from five point differences of the actual computed profiles, using $\sigma$ and $m$ to retain small lapse and metric corrections. For an expanded equation $E_i=\sum_j E_{ij}$, define
\begin{equation}
 \epsilon_{\rm EL}=\max_{i,r}\frac{|E_i|}{\sum_j|E_{ij}|}.
 \label{eq:finiteELresidual}
\end{equation}
The difference test uses $r/r_h=1.2,2,4,10$ and successive steps $\Delta t=0.012,0.006,0.003$. Its first halving resolves the fourth order difference error; the smallest step gives $\epsilon_{\rm EL}<1.8\times10^{-10}$ for all four backgrounds. These comparisons support the numerical solutions and do not constitute interval error bounds or an analytic global existence theorem.

The weak deformation limit also checks the connection to Section~\ref{sec:maxwell}. At $r=4M$, the ratio $X_\lambda/(\eta\chi_0)$ decreases from $1.06915$ at $\lambda=0.1$ to $1.00552$ at $\lambda=0.003$, while $(N_\lambda-1)/(4\alpha\eta\chi_0)$ decreases from $1.06134$ to $1.00527$. Here $\chi_0$ is the independently derived response in Eq.~\eqref{eq:closedchi} on the $Q/M=1/2$ seed. The agreement concerns the norm and lapse responses; the homogeneous hair convention is the one in Eq.~\eqref{eq:farlengthconvention}.

\begin{figure}[htbp]
\centering
\includegraphics[width=\textwidth]{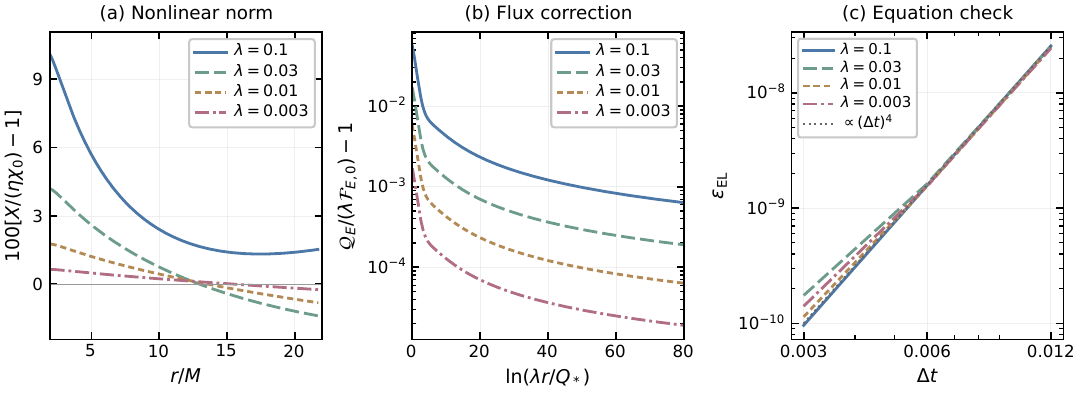}
\caption{Full finite deformation exteriors at $\alpha/M^2=0.1$ and $Q_*/M=1/2$. (a) Percentage departure of the full norm from $\eta\chi_0$ in the near exterior. Each curve starts at its own horizon; $\chi_0$ is the null seed response with $Q/M=1/2$. (b) The positive flux excess $\mathcal Q_E/(\lambda\mathcal F_{E,0})-1$ for $\ln(\lambda r/Q_*)\ge0$. Here $\mathcal F_{E,0}=2k_0^2/q_0$ is the leading outer connection of Section~\ref{sec:far}, evaluated at the same $\rho$ and normalized by Eq.~\eqref{eq:radiusnormalization}. The logarithmic ordinate resolves its finite coupling correction. (c) Residuals of all four original regular Euler equations using five point profile derivatives, with a fourth order guide. All full backgrounds use Eq.~\eqref{eq:farlengthconvention}.}
\label{fig:finitebackgrounds}
\end{figure}

Figure~\ref{fig:finitebackgrounds} resolves the relative norm correction in the near exterior, the nonlinear far flux connection, and the original equation difference test. The backgrounds satisfy the two local radial principal conditions of Eq.~\eqref{eq:commonframeconditions} at every sampled exterior point. Their necessary even angular test is evaluated in Section~\ref{sec:angular}; it is logically independent of the background equations and these two radial signs.

\section{Longitudinal mode and the angular obstruction}\label{sec:angular}
\subsection{The vector constraint at principal order}
Let $\delta W_\mu=B_\mu+\nabla_\mu\pi$, with dimensionless $\pi$. This parametrization has the redundancy $B_\mu\mapsto B_\mu+\nabla_\mu\chi$, $\pi\mapsto\pi-\chi$; it adds no physical mode. Write $S_{\mu\nu}=\nabla_{(\mu}W_{\nu)}$ and $\Theta=\nabla_\mu W^\mu$. Expansion of the cubic interaction followed by integration by parts gives
\begin{equation}
 \mathcal L^{(2)}_{3,\pi}=16\alpha(S^{\mu\nu}-\Theta g^{\mu\nu})
                    \nabla_\mu\pi\nabla_\nu\pi.
 \label{eq:scalarcubic}
\end{equation}
Background derivatives must be retained before this integration. The two derivative mixing with $B_\mu$ is $16\alpha F^{\mu\nu}[B]W_\nu\nabla_\mu\pi$. Completing the Maxwell square with
\begin{equation}
 \widehat B_\mu=B_\mu-\frac{16\alpha}{\eta}W_\mu\pi
 \label{eq:scalarvectorshift}
\end{equation}
adds $(128\alpha^2/\eta)(W^2g^{\mu\nu}-W^\mu W^\nu)$ to the scalar tensor. Including the curvature and quartic terms, the fixed metric vector Schur tensor is
\begin{align}
 Z^{\mu\nu}={}&16\alpha(S^{\mu\nu}-\Theta g^{\mu\nu})-4\alpha G^{\mu\nu}
       -12\alpha W^2g^{\mu\nu}-24\alpha W^\mu W^\nu\nonumber\\
 &+\frac{128\alpha^2}{\eta}(W^2g^{\mu\nu}-W^\mu W^\nu).
 \label{eq:scalarZ}
\end{align}
The quadratic scalar action contains $Z^{\mu\nu}\nabla_\mu\pi\nabla_\nu\pi$, with the same positive overall normalization as before. Independently, the two derivative part of $(16\alpha/\eta)W_\mu\delta E^\mu-\nabla_\mu\delta E^\mu$, using the original vector equation, gives the Euler derivative of Eq.~\eqref{eq:scalarZ}. This checks the constraint contribution without freezing the vector perturbation.

\subsection{The coupled even angular system}
We retain the metric perturbation in the reduction. In a local orthonormal frame let $x$ be radial and let the wave covector be $k_\mu=(-\upsilon,0,1,0)$, after division by the angular wave number. Define $w_0=A/\sqrt f$, $w_1=g/\sqrt f$. The even angular gauges $h_{0y}=h_{xy}=h_{yy}=0$ and $\widehat B_y=0$ leave seven amplitudes,
\begin{equation}
 \boldsymbol X=(U,V,\Sigma,\mathcal B_0;\Gamma,\mathcal B_1,\pi),\qquad
 U=h_{00},\quad V=h_{0x},\quad
 \Sigma=h_{xx}+h_{zz},\quad\Gamma=h_{xx}-h_{zz}.
 \label{eq:angularfields}
\end{equation}
The complete local principal quadratic form is given in Appendix~\ref{app:angular}. In particular its metric scalar mixing is $8\alpha\pi h^{\mu\nu}\mathsf E_{k,\mu\nu}[T]$, where $\mathsf E_k$ is the linearized Einstein symbol and
\begin{equation}
 T_{\mu\nu}=S_{\mu\nu}+W^2g_{\mu\nu}
                 -\left(2+\frac{16\alpha}{\eta}\right)W_\mu W_\nu.
 \label{eq:scalarmetricmix}
\end{equation}
The curvature expansion agrees with the Einstein symbol and has the four metric gauge null directions before gauge fixing.

Write the quadratic form as $\boldsymbol X^{\mathsf T}\boldsymbol M\boldsymbol X$. For the first four amplitudes its constraint block satisfies
\begin{equation}
 \det\boldsymbol M_{CC}=-\frac{(1+2\alpha w_0^2)^2}{64}
       \left[\eta\{1+2\alpha(w_0^2-w_1^2)\}-16\alpha^2w_1^2\right].
 \label{eq:angularconstraintdet}
\end{equation}
Where this is nonzero, elimination gives a symmetric three mode pencil,
\begin{equation}
 \boldsymbol M_{\rm phys}
 =\boldsymbol M_{PP}-\boldsymbol M_{PC}\boldsymbol M_{CC}^{-1}\boldsymbol M_{CP}
 =\upsilon^2\boldsymbol K_\Omega-\boldsymbol P_\Omega,
 \qquad \det(\boldsymbol P_\Omega-c_\Omega^2\boldsymbol K_\Omega)=0.
 \label{eq:angularpencil}
\end{equation}
Its entries are affine in $\upsilon^2$; the eliminated fields introduce no extra frequency root. We use this gauge only in its nonsingular far region. Treating the three interaction coefficients independently and removing the nonminimal and quartic interactions recovers the cubic theory angular speed of Ref.~\cite{DeFelice2024}, as shown in the appendix. The angular calculation thus retains both the gravitational constraints and the Maxwell Schur term.

\subsection{The first nonzero logarithmic angular coefficient}
In the flat far balance, use $a=\lambda q/r$, $g=\lambda k/r$ and dots for derivatives with respect to $\ln r$. With $Z=(16\alpha\lambda/r^2)\overline Z$, Eq.~\eqref{eq:scalarZ} gives
\begin{align}
 \overline Z^{00}&=\dot k+k+\frac{3\lambda}{4}(k^2-3q^2)-8k^2,\nonumber\\
 \overline Z^{01}&=-\frac{\dot q-q}{2}+\left(8+\frac{3\lambda}{2}\right)qk,\nonumber\\
 \overline Z^{11}&=-2k+\frac{3\lambda}{4}(q^2-3k^2)-8q^2,\nonumber\\
 \overline Z^{\Omega}&=-\dot k+\left(\frac{3\lambda}{4}-8\right)(q^2-k^2).
 \label{eq:scalarfarcomponents}
\end{align}
The finite $\lambda$ slow branch of Eq.~\eqref{eq:finitefarbalance} has $k=q/\sqrt2+(4+3\lambda/16)q^2+O(q^3)$, $\dot q=-8\sqrt2q^2+O(q^3)$ and $\dot k=-8q^2+O(q^3)$. Hence
\begin{equation}
 \overline Z^{00}=\frac q{\sqrt2}+O(q^2),\qquad
 \overline Z^\Omega=\left(4+\frac{3\lambda}{8}\right)q^2+O(q^3).
 \label{eq:scalarfarsigns}
\end{equation}
The order $q$ angular term cancels. Omitting the Maxwell constraint would replace the coefficient $4+3\lambda/8$ by $8+3\lambda/8$, so it cannot be neglected at this order.

Metric mixing does not remove this sign. Equation~\eqref{eq:scalarmetricmix} is of order $\alpha(\nabla W+W^2+\alpha W^2/\eta)$ in the quadratic action. Its Einstein Schur correction is of order $\alpha^2(\nabla W+W^2+\alpha W^2/\eta)^2$, with relative order $\alpha\lambda/r^2$ compared with the first nonzero angular coefficient at fixed $\lambda>0$. The complete matrix in Eq.~\eqref{eq:angularpencil}, after rescaling $\pi$ by $r$, independently recovers Eq.~\eqref{eq:scalarfarcomponents} at leading inverse radius order. Corrections from the asymptotically flat metric jets are also smaller than every retained inverse logarithm. The longitudinal root is separated from the other two angular roots, which approach unity. It follows that the physical coupled root has
\begin{align}
 c_{\Omega,L}^2&=-\sqrt2\left(4+\frac{3\lambda}{8}\right)q+O(q^2)
                      +O\!\left(\frac Mr\right)+O\!\left(\frac{\alpha\lambda q}{r^2}\right)\nonumber\\
 &=-\frac{\tfrac12+\tfrac{3\lambda}{64}}{\ln(r/r_*)}
    +O\!\left(\frac{\ln\ln(r/r_*)}{\ln^2(r/r_*)}\right)<0
 \label{eq:angularobstruction}
\end{align}
at sufficiently large finite radius on the positive logarithmic branch. This is a failure of the necessary angular condition for $\alpha>0$, $\eta>0$. The tensor and transverse vector radial conditions in Eq.~\eqref{eq:commonframeconditions} remain compatible with this failure.

The radial scalar block has $\overline Z^{01}=q/2+O(q^2)$, $\overline Z^{11}=-\sqrt2q+O(q^2)$ and determinant $-5q^2/4+O(q^3)$. Its limiting radial roots $(1\pm\sqrt5)/\sqrt2$ include the mixed term. More explicitly, completing the time square gives the spatial Schur form
\begin{equation}
 \overline Z^{ij}-\frac{\overline Z^{0i}\overline Z^{0j}}{\overline Z^{00}}
 =\operatorname{diag}\!\left(-\frac{5\sqrt2}{4}q,
             \left(4+\frac{3\lambda}{8}\right)q^2,
             \left(4+\frac{3\lambda}{8}\right)q^2\right)
   +\operatorname{diag}\big(O(q^2),O(q^3),O(q^3)\big).
 \label{eq:scalarspatialSchur}
\end{equation}
The far scalar tensor therefore has inertia $(3+,1-)$. An invertible frame change preserves this inertia. With a positive time coefficient, the spatial Schur form retains two positive eigenvalues; positive scalar kinetic and spatial energies would instead require inertia $(1+,3-)$ for the full principal tensor. A radial boost also leaves the angular entries unchanged. The suppressed metric correction discussed above cannot reverse their sufficiently far sign. This states the frame content of the obstruction within the controlled far principal reduction.

For the far fixed metric scalar polynomial displayed below, put $\mu=|k_r|/|\boldsymbol k|$, $a_\Omega=Z^\Omega/Z^{00}>0$ and $D_r=[(Z^{01})^2-Z^{00}Z^{11}]/(Z^{00})^2$. Its imaginary frequency and angular band obey
\begin{equation}
 \frac{\operatorname{Im}\widehat\omega}{|\boldsymbol k|}
 =\sqrt{[a_\Omega(1-\mu^2)-D_r\mu^2]_+},\qquad
 \mu_{\rm crit}^2=\frac{a_\Omega}{a_\Omega+D_r}
 \sim\frac{\tfrac15+\tfrac{3\lambda}{160}}{\ln(r/r_*)},
 \label{eq:directionalband}
\end{equation}
where $[x]_+=\max(x,0)$ and $D_r\to5/2$. This relation describes the narrowing local angular band of that far approximation.

\subsection{Evaluation on the finite deformation backgrounds}
We evaluate Eq.~\eqref{eq:angularpencil} on the full backgrounds of Table~\ref{tab:finitehorizons}, using covariant orthonormal derivatives and the complete Einstein tensor. Each coupling has $1000$ samples over $5\leq\ln\rho\leq160$. All $4000$ longitudinal squared angular speeds are negative. After positive diagonal normalization, the time kinetic matrices have minimum eigenvalue above $0.99985$. This tests their positive inertia at each sampled finite point, without bounding the absolute kinetic scale. The unnormalized longitudinal time coefficient scales as $(16\alpha\lambda/r^2)q/\sqrt2$ in the far limit. The normalized constraint margin in Eq.~\eqref{eq:angularconstraintdet} exceeds $0.9999999992$, and the other two angular roots are positive.

\begin{table}[htbp]
\centering\small
\begin{tabular}{rrrrr}
\toprule
$\lambda$ & $c_{\Omega,L}^2(\ln\rho=10)$ & $c_{\Omega,L}^2(40)$ & $c_{\Omega,L}^2(160)$ & $\tfrac12+3\lambda/64$\\\midrule
$0.1$ & $-0.110729338$ & $-0.0162126903$ & $-0.00342168844$ & $0.504687500$\\
$0.03$ & $-0.109121735$ & $-0.0160831389$ & $-0.00339816387$ & $0.501406250$\\
$0.01$ & $-0.10866654$ & $-0.0160461825$ & $-0.00339144526$ & $0.500468750$\\
$0.003$ & $-0.10850765$ & $-0.0160332539$ & $-0.00338909403$ & $0.500140625$\\\bottomrule
\end{tabular}
\caption{Physical longitudinal angular root of the complete constrained even system at $\alpha/M^2=0.1$ and $Q_*/M=1/2$. The last column is the exact asymptotic coefficient, not a fitted value. The slow approach includes the logarithm of the logarithm in Eq.~\eqref{eq:angularobstruction}.}
\label{tab:angularmodes}
\end{table}

Table~\ref{tab:angularmodes} gives representative physical roots. Replacing the primary background integration by the independent implicit integration changes these roots by less than $6.9\times10^{-11}$ relatively at the comparison radii. Changing the far starting amplitude or slow series order at fixed far length changes the roots by less than $8.9\times10^{-10}$ relatively. Five point profile derivatives give relative changes below $1.7\times10^{-11}$ at the final difference step. A separate evaluation of the determinant at $90$ decimal digits agrees within $6.7\times10^{-15}$ relatively for the same input jets. This comparison isolates matrix roundoff from errors in the background and its derivatives. The independent temporal gauge reduction in Appendix~\ref{app:angular} reproduces all three physical roots within $1.1\times10^{-13}$ relatively at the $24$ comparison points. Reconstruction satisfies all eleven unfixed principal equations, with maximum normwise polynomial backward error below $5.9\times10^{-16}$ in normalized field units. These backward errors measure the matrix solves, separately from background accuracy.

\begin{figure}[htbp]
\centering
\includegraphics[width=\textwidth]{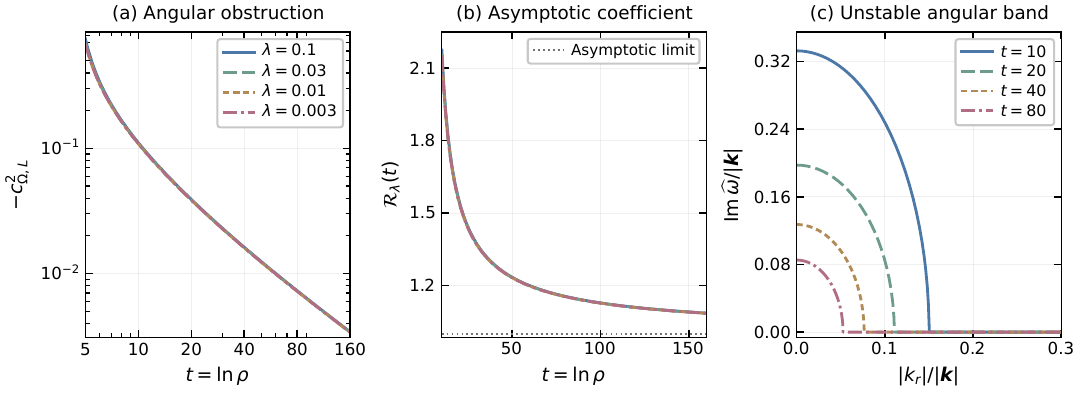}
\caption{Longitudinal angular obstruction at $\alpha/M^2=0.1$ and $Q_*/M=1/2$. (a) The negative physical root, displayed as $-c_{\Omega,L}^2>0$ with logarithmic axes. (b) The normalized coefficient $\mathcal R_\lambda(t)$ approaches unity, with the same coupling colors and line styles as (a). The close curves reflect the weak coupling dependence. (c) Local imaginary frequency for $\lambda=0.1$ at four radii, from the far scalar polynomial \eqref{eq:directionalband}. The curves vanish outside its unstable angular band. Panels (a) and (b) use the full constrained physical pencil; panel (c) is the stated far longitudinal approximation.}
\label{fig:angularobstruction}
\end{figure}

Figure~\ref{fig:angularobstruction} shows the negative physical angular root and its approach to the analytic coefficient. Define $\mathcal R_\lambda(t)=-t c_{\Omega,L}^2/(1/2+3\lambda/64)$, where $t=\ln\rho$; its limit is unity. At $t=160$ it is $1.08477$ for $\lambda=0.1$, consistent with the slow subleading logarithms.  The third panel displays Eq.~\eqref{eq:directionalband}. The finite exteriors pass the radial tensor and transverse vector tests while failing the physical even angular test.

\subsection{Physical interpretation}
The radial and angular tests probe distinct physical directions. Positive radial tensor and transverse vector forms do not constrain the longitudinal angular gradient energy. The negative root in Eq.~\eqref{eq:angularobstruction} belongs to the reduced physical system and survives the independent gauge calculation.

For a local angular packet, $\widehat\omega^2\simeq c_{\Omega,L}^2\ell(\ell+1)/r^2$. Take the short wavelength limit at a fixed finite far annulus, with sufficiently large $\ell$, before sending its radius to infinity. The vanishing time coefficient at infinity supplies neither a global black hole eigenfrequency nor an effective cutoff.

\section{Conclusion}\label{sec:conclusion}
The original nontrivial constant norm family cannot have both positive dipole kinetic energy and real nondegenerate radial tensor characteristics for the specified action. At its degenerate boundaries, the regular canonical reduction retains one physical dipole pair and the tensor propagator exhibits derivative loss.

A Maxwell term for the same vector excludes nontrivial constant norm radial hair under the stated asymptotic and nondegeneracy assumptions. The varying norm response is nonuniform at large radius. Its positive logarithmic continuation has finite ADM mass and numerical exteriors connecting to simple future horizons, supported by the exact ingoing reduction and independent equation checks.

The continuation nevertheless fails a local angular admissibility condition: the constrained even system has a longitudinal root with negative squared speed at sufficiently large finite radius. Both the Maxwell constraint and quartic interaction contribute to its inverse logarithm coefficient in Eq.~\eqref{eq:angularobstruction}. Independent gauges and four full backgrounds confirm this sign where the time kinetic form is positive. The resulting obstruction is specific to the action and branches examined here.

\appendix
\section{Variation identities and spherical completeness}\label{app:variation}
Varying the three vector terms in Eq.~\eqref{eq:LV} gives, before the common factor $-8\alpha$ is removed,
\begin{align}
 \delta_W(4G^{\rho\sigma}W_\rho W_\sigma)&=8G^{\mu\nu}W_\nu\delta W_\mu,\\
 \delta_W(8W^2\nabla_\rho W^\rho)&=
 \left[16W^\mu\nabla_\rho W^\rho-8\nabla^\mu W^2\right]\delta W_\mu,
\end{align}
up to a boundary derivative. The quartic term contributes $24W^2W^\mu\delta W_\mu$. These expressions give Eq.~\eqref{eq:vectorE}.

For the background variation, a static off diagonal metric can be included as
\begin{equation}
 \dd s^2=-N^2f(\dd t+s(r)\dd r)^2+\frac{\dd r^2}{f}+r^2\dd\Omega^2.
\end{equation}
The local change $\dd T=\dd t+s(r)\dd r$ leaves the radial vector component equal to $b-as$ in the diagonal coordinates. Consequently the shift variation is $E_s=-aE_b$. It vanishes on the independently varied vector equation. The remaining angular metric equation follows from the diffeomorphism identity on a regular spherical interval once the other metric and vector equations hold.

Differentiating the first integral gives
\begin{equation}
 u'=-\frac{2u(r+\alpha u)}{r(r+4\alpha u)}.
\end{equation}
Differentiating once more yields the second form of $V_1$ in Eq.~\eqref{eq:Vdip}. At a regular point on the displayed square root branch, $r+4\alpha u=r\sqrt{1+8\alpha Q/r^3}\ne0$.

\section{Generalized Proca convention check}\label{app:dictionary}
The identity
\begin{align}
 &(\nabla_\mu W^\mu)^2-\nabla_\mu W_\nu\nabla^\nu W^\mu\nonumber\\
 &\hspace{8mm}=R_{\mu\nu}W^\mu W^\nu+
 \nabla_\mu\left[W^\mu\nabla_\nu W^\nu-W^\nu\nabla_\nu W^\mu\right]
\end{align}
shows that $G_4=1-4\alpha X$ reproduces $R-4\alpha G^{\mu\nu}W_\mu W_\nu$ after integration by parts. The cubic term is $16\alpha X\nabla_\mu W^\mu$ with the convention used in Eq.~\eqref{eq:dictionary}.

The action convention in Ref.~\cite{Kase2018} contains an explicit Maxwell scalar $\mathcal F=-W_{\mu\nu}W^{\mu\nu}/4$ in addition to $G_2(X,\mathcal F)$. In the unfactored reference action, dimensional normalization requires $G_4^{\rm ref}=p(1-4\alpha X)$ and $G_3^{\rm ref}=16p\alpha X$, where $p=\mpl^2/2$. To represent Eq.~\eqref{eq:action}, set
\begin{equation}
 G_2^{\rm ref}(X,\mathcal F)=-\mathcal F-24p\alpha X^2,
 \qquad 1+G^{\rm ref}_{2,\mathcal F}=0.
\end{equation}
This is why $C_5=0$ in Eq.~\eqref{eq:Ccheck}. It also fixes a possible ambiguity in applying a formula that was written with a canonical Maxwell term separated from the rest of the action.

For the kinetic deformation \eqref{eq:actioneta}, the same unfactored convention instead requires
\begin{equation}
 G_{2,\eta}^{\rm ref}=(p\eta-1)\mathcal F-24p\alpha X^2,
 \qquad 1+G_{2,\eta,\mathcal F}^{\rm ref}=p\eta.
\end{equation}
The product $p\eta$ is dimensionless. Dividing the spherical coefficients by $p$ then reproduces the direct principal coefficient \eqref{eq:staticKeta}.

\section{Complete angular principal form and constraints}\label{app:angular}
All contractions in this appendix use $\eta_{\mu\nu}=\operatorname{diag}(-1,1,1,1)$, $W_\mu=(w_0,w_1,0,0)$ and $k_\mu=(-\upsilon,0,1,0)$. For a symmetric tensor $H_{\mu\nu}$, the linearized Einstein symbol is
\begin{equation}
 \mathsf E_{k,\mu\nu}[H]=\frac12\left[
 -k_\mu k^\rho H_{\rho\nu}-k_\nu k^\rho H_{\rho\mu}
 +k^2H_{\mu\nu}+k_\mu k_\nu H
 +\eta_{\mu\nu}(k^\rho k^\sigma H_{\rho\sigma}-k^2H)\right],
 \label{eq:Einsteinsymbol}
\end{equation}
where $H=\eta^{\mu\nu}H_{\mu\nu}$ and $k^\mu\mathsf E_{k,\mu\nu}=0$. In the gauge of Eq.~\eqref{eq:angularfields}, the direct curvature expansions give
\begin{align}
 \mathscr L_{\rm EH}&=\frac{\upsilon^2-1}{8}(\Gamma^2-\Sigma^2)
                         -\frac{\Sigma U}{2}+\frac{V^2}{2},\nonumber\\
 16\mathcal Q_k&=\mathcal A_k(\Sigma^2-\Gamma^2)
       +4w_0^2\Sigma U-4w_1^2\Gamma U
       +4(w_1^2-w_0^2)V^2+8w_0w_1 V(\Gamma-\Sigma),\nonumber\\
 \mathcal A_k&=w_0^2(1+\upsilon^2)+w_1^2(1-\upsilon^2).
 \label{eq:angularcurvaturepolynomial}
\end{align}
Here $\mathcal Q_k$ is the quadratic coefficient of $\sqrt{-g}G^{\mu\nu}W_\mu W_\nu$, with the covariant background vector held fixed. Retaining the first background jet for the mixed terms, the complete local principal amplitude action is
\begin{align}
 \mathscr L_k={}&\mathscr L_{\rm EH}-4\alpha\mathcal Q_k
       +\frac\eta2\left[\mathcal B_0^2+(\upsilon^2-1)\mathcal B_1^2\right]
       -8\alpha W^\mu\widehat B^\nu\mathsf E_{k,\mu\nu}[h]\nonumber\\
 &+8\alpha\pi h^{\mu\nu}\mathsf E_{k,\mu\nu}[T]
       +\pi^2 Z^{\mu\nu}k_\mu k_\nu.
 \label{eq:completeangularaction}
\end{align}
Equations~\eqref{eq:scalarZ}, \eqref{eq:scalarmetricmix}, \eqref{eq:angularcurvaturepolynomial} and \eqref{eq:completeangularaction} specify every entry of $\boldsymbol M=\tfrac12\partial^2\mathscr L_k/\partial\boldsymbol X\partial\boldsymbol X^{\mathsf T}$. The lapse equation fixes $\Sigma$, the shift and electric equations form a two dimensional block, and the remaining trace equation fixes $U$. Direct reconstruction satisfies all four constraint rows. This ordering gives Eq.~\eqref{eq:angularpencil} without a large expanded determinant.

We independently expand the unfixed even action with eleven amplitudes and impose the alternative gauge $h_{00}=h_{0x}=h_{0y}=\widehat B_0=0$. Put $J_y=h_{xy}$, $D_y=h_{yy}$ and $\boldsymbol C_{\rm tm}=(J_y,D_y,\Sigma,\mathcal B_y)^{\mathsf T}$. The leading gauge transformation to the spatial choice is
\begin{equation}
 \boldsymbol C_{\rm sp}=(\upsilon^2D_y,\upsilon J_y,\Sigma,\upsilon\mathcal B_y)^{\mathsf T}
       =\mathsf A(\upsilon)\boldsymbol C_{\rm tm},\qquad
 \det\mathsf A=-\upsilon^4.
 \label{eq:gaugetransformation}
\end{equation}
The separately restricted actions satisfy $\boldsymbol M_{\rm tm}=\mathsf R^{\mathsf T}\boldsymbol M_{\rm sp}\mathsf R$, with $\mathsf R=\operatorname{diag}(\mathsf A,\boldsymbol I_3)$. Direct elimination in the temporal gauge verifies
\begin{equation}
 \det\boldsymbol M_{CC,\rm tm}=\upsilon^8\det\boldsymbol M_{CC,\rm sp},\qquad
 \boldsymbol M_{{\rm phys},\rm tm}=\boldsymbol M_{{\rm phys},\rm sp}
 \quad(\upsilon\ne0).
 \label{eq:gaugeidentity}
\end{equation}
All four principal gauge identities hold in the eleven row system. Reconstructing the temporal gauge solution annihilates its eight nonphysical rows before selecting a characteristic root, so the three physical equations imply every unfixed equation. The comparison applies to the nonzero angular roots; it does not divide by zero frequency to classify a static mode.

For the static background, let $\nu=N'/N$. The covariant jets supplied to this action are
\begin{align}
 S_{00}&=-g\left(\nu+\frac{f'}{2f}\right),&
 S_{01}&=\frac12\left(A'-A\nu-\frac{Af'}f\right),\nonumber\\
 S_{11}&=g'-\frac{gf'}{2f},&S_{22}=S_{33}&=\frac gr,\nonumber\\
 G_{00}&=\frac{1-f-rf'}{r^2},&
 G_{11}&=\frac{f-1+rf'+2rf\nu}{r^2},\nonumber\\
 G_{22}=G_{33}&=f\frac{N''}N+\frac32f'\nu+\frac{f''}2
                           +\frac{f\nu}r+\frac{f'}r.
 \label{eq:angularsphericaljets}
\end{align}
These are covariant orthonormal entries; the time radial entry changes sign when both indices are raised in Eq.~\eqref{eq:scalarZ}. For numerical evaluation, Eq.~\eqref{eq:finiteflow} also gives $r^2G_{22}=-\dot S/2+f\dot\sigma^2+\dot f\dot\sigma/2$, where $S$ is defined in Eq.~\eqref{eq:finiteVS} and dots denote derivatives with respect to $\ln(r/M)$. This form avoids cancellation of the vacuum curvature terms. A separate spherical Christoffel calculation verifies all entries in Eq.~\eqref{eq:angularsphericaljets}.

As an external calibration, remove the Einstein tensor coupling and the quartic potential while retaining the cubic interaction. On its constant temporal far background, the reduction gives two luminal modes and
\begin{equation}
 c_{\Omega,\rm cubic}^2=\frac{4-\eta w_0^2}{3\eta w_0^2}.
 \label{eq:cubiccalibration}
\end{equation}
Canonical normalization $A_0=\sqrt\eta w_0$ and the coefficient $1$ of $R$ correspond to $M_{\rm Pl}^2=2$ in Ref.~\cite{DeFelice2024}; Eq.~\eqref{eq:cubiccalibration} then reproduces its asymptotic result. The present logarithmic branch instead has $A_0\to0$, and its speed must be obtained from the next nonvanishing terms in Eq.~\eqref{eq:scalarfarsigns}.

\section{Regular constraint reduction at the axial boundaries}\label{app:boundaries}
\subsection{The complete constant norm dipole action}
The inverse factor $H^{-1}$ in Eq.~\eqref{eq:Kdip} results from eliminating a rotational variable. It must not be used to count constraints at $H=0$. Keeping both $P$ and $\Omega$ instead gives a regular description. The complete dipole action, after use of the background relations and removal of boundary derivatives, is
\begin{align}
 L_1^{(2)}&=\frac q2\mathcal F_1^2+
  \mathcal F_1\big[B\dot a_1+S[a_1]\big]+V_1a_1^2,
  \label{eq:completegeneral}\\
 q&=r^4H,\qquad B=\frac{4\alpha r^2g}{f},
  \label{eq:qB}\\
 S[a_1]&=-4\alpha r^2\left[Aa_1'+\left(A'-\frac{2A}r\right)a_1\right].
  \label{eq:Soperator}
\end{align}
The positive factors removed in Section~\ref{sec:dipole} remain understood. Direct Euler differentiation of the covariantly expanded action and Eq.~\eqref{eq:completegeneral} gives identical equations for $P$, $\Omega$, and $a_1$, for either smooth sign of $A$. This comparison checks the terms without derivatives as well as the principal terms.

The potential is independent of the norm constant:
\begin{align}
 V_1&=4\alpha\left(ru''+2u'-\frac{2u}r\right)
 =-\frac{144\alpha^2u^2(r+2\alpha u)^2}{r(r+4\alpha u)^3}.
 \label{eq:generalpotential}
\end{align}
For example, before substitution of $g=u+cr$ and the metric relation, the algebraic coefficient is
\begin{equation}
 24\alpha c-\frac{16\alpha g}r-\frac{4\alpha f'}r
 -2\alpha f''-8\alpha g'.
\end{equation}
All terms proportional to $c$ cancel. For $c=0$, Eq.~\eqref{eq:completegeneral} reduces to the null action derived above. The general dipole structure is consistent with Ref.~\cite{Kase2018}, but here its full specialization follows directly from the original vector action.

\subsection{A regular Legendre map and the local constraint count}
Let $\pi_P$ and $\pi_1$ denote the momenta conjugate to $P$ and $a_1$. Equation~\eqref{eq:completegeneral} gives
\begin{align}
 \pi_P&=q\mathcal F_1+B\dot a_1+S[a_1],&
 \pi_1&=B\mathcal F_1,& \pi_\Omega&=0.
 \label{eq:boundarymomenta}
\end{align}
The velocity Hessian for $(P,a_1)$ is
\begin{equation}
 \mathbb W=\begin{pmatrix}q&B\\B&0\end{pmatrix},
 \qquad \det\mathbb W=-B^2.
 \label{eq:velocityhessian}
\end{equation}
Thus $q=0$ does not introduce a primary dipole constraint when $B\ne0$. The inverse Legendre map is
\begin{equation}
 \dot P=\Omega'+\frac{\pi_1}{B},\qquad
 \dot a_1=\frac{\pi_P-S[a_1]}B-\frac{q\pi_1}{B^2}.
 \label{eq:inverselegendre}
\end{equation}
No division by $q$ or $H$ occurs.

The canonical Hamiltonian density, before the rotational constraint is imposed, is
\begin{equation}
 \mathcal H_1=\pi_P\Omega'
 +\frac{\pi_1(\pi_P-S[a_1])}{B}
 -\frac{q\pi_1^2}{2B^2}-V_1a_1^2.
 \label{eq:fullHamiltonian}
\end{equation}
Preservation of $\pi_\Omega=0$ gives $\pi_P'=0$. Since $P$ is absent from the Hamiltonian, $\dot\pi_P=0$ as well. The two local constraints $\pi_\Omega=0$ and $\pi_P'=0$ are first class and close without a further constraint. On a connected interval their remaining integration constant is the homogeneous angular momentum $j$. It is fixed to zero for the same rotational sector used above. After fixing the axial coordinate freedom, the local count is
\begin{equation}
 n_{\rm dipole}=\frac{6-2\times2}{2}=1,\qquad B\ne0.
 \label{eq:dipolecount}
\end{equation}
The global rotational constant is not included in this count. The reduced symplectic pair is $(a_1,\pi_1)$, and its Hamiltonian is
\begin{align}
 \mathcal H_{1,\mathrm{red}}&=
 \pi_1\mathscr D a_1-\frac{q\pi_1^2}{2B^2}-V_1a_1^2,
 \label{eq:regularHred}\\
 \mathscr D a_1&\equiv-\frac{S[a_1]}B
 =\frac f g\left[Aa_1'+\left(A'-\frac{2A}r\right)a_1\right].
 \label{eq:Dgeneral}
\end{align}
For $q\ne0$, its momentum square coefficient equals $1/(4\Kt)$, recovering Eq.~\eqref{eq:Kdip}. Equations~\eqref{eq:regularHred} and \eqref{eq:dipolecount} remain valid at $H=0$.

\subsection{The critical dipole Hamiltonian}
At $\zeta=-1$, $q=0$. Equation~\eqref{eq:regularHred} becomes
\begin{equation}
 \mathcal H_{1,-}=\pi_1\mathscr D a_1-V_1a_1^2.
 \label{eq:criticalH}
\end{equation}
The absence of a momentum square is accompanied by a surviving physical canonical pair. Its exact quadratic equations are
\begin{align}
 \dot a_1&=v_1a_1'+w_1a_1,\\
 \dot\pi_1&=(v_1\pi_1)'-w_1\pi_1+2V_1a_1,
 \label{eq:criticalevolution}\\
 v_1&=\frac{fA}{g},\qquad w_1=\frac f g\left(A'-\frac{2A}r\right).
\end{align}
They do not impose $\pi_1=0$ or $a_1=0$.

On a regular static interval, choose any smooth compactly supported $a_1$ with $\mathscr D a_1\not\equiv0$. For a positive reference length $L_{\rm ref}$, the allowed initial momenta
\begin{equation}
 \pi_1=-\Lambda_E L_{\rm ref}^2\mathscr D a_1,\qquad \Lambda_E>0,
 \label{eq:negativepacket}
\end{equation}
give the integrated static canonical energy, with the positive common normalization suppressed,
\begin{equation}
 E_{1,-}=-\Lambda_E L_{\rm ref}^2\int\dd r\,(\mathscr D a_1)^2
              -\int\dd r\,V_1a_1^2.
 \label{eq:unboundedenergy}
\end{equation}
The potential contribution is finite, and the first term makes the quadratic form unbounded below. A negative direction can be chosen at finite $\Lambda_E$ and then scaled to arbitrarily small perturbation amplitude. Compact support removes endpoint improvements to the integrated canonical energy. This is a statement about the physical static Hamiltonian, not a computed exponential instability rate.

Such a test field exists on every open region with $A\not\equiv0$. The background first integral at $c=-1/(4\alpha)$ gives the exact identity
\begin{equation}
 A^2=\frac{r^2}{16\alpha^2}-\frac1{2\alpha}
       +\frac{M-Q/2}{\alpha r}.
 \label{eq:criticalA}
\end{equation}
Its cubic numerator cannot vanish on an open interval. Isolated zeros do not obstruct choosing a smaller regular support. At the other boundary, $\zeta=1$, $H=2$ and the nonsingular reduction already gives
\begin{equation}
 \Kt=-\frac{4\alpha^2g^2}{f^2}<0.
 \label{eq:plusboundaryK}
\end{equation}

\subsection{Tensor rank and the repeated characteristic}
The tensor principal action can also be evaluated at the boundaries directly, without a characteristic discriminant test. In local orthonormal coordinates $(\tau,x)$, Eq.~\eqref{eq:TT} becomes
\begin{align}
 L_{T,-}^{(2)}&=2\alpha(\mathsf B\,\partial_\tau h
                              -\mathsf A\,\partial_xh)^2,
             &&\zeta=-1,\label{eq:tensorminus}\\
 L_{T,+}^{(2)}&=2\alpha(\mathsf A\,\partial_\tau h
                              -\mathsf B\,\partial_xh)^2,
             &&\zeta=1.\label{eq:tensorplus}
\end{align}
Both principal quadratic forms have rank one. An invertible change of coordinates preserves that rank, so neither defines a Lorentzian tensor cone. If the coefficient of $\partial_\tau h$ is nonzero, divide by its finite constant value in the frozen equation and define
\begin{equation}
 v_-={\mathsf A}/{\mathsf B},\qquad
 v_+={\mathsf B}/{\mathsf A},\qquad
 (\partial_\tau-v_\pm\partial_x)^2h=0.
 \label{eq:squaredtransport}
\end{equation}
Here $v_\pm$ are transport coefficients; the repeated characteristic ray has $\dd x/\dd\tau=-v_\pm$. If the time coefficient instead vanishes, the chosen time surface is characteristic and does not give an ordinary tensor evolution equation.

Real repetition alone is insufficient for a well behaved wave Cauchy problem. The derivative variables
\begin{equation}
 U=\begin{pmatrix}\partial_xh\\
             (\partial_\tau-v_\pm\partial_x)h\end{pmatrix}
\end{equation}
obey the frozen principal system
\begin{equation}
 \partial_\tau U=\mathbb A_\pm\partial_xU,\qquad
 \mathbb A_\pm=\begin{pmatrix}v_\pm&1\\0&v_\pm\end{pmatrix}.
 \label{eq:tensorJordan}
\end{equation}
Its nontrivial Jordan block prevents a frequency independent estimate in the standard norm of the first derivatives, which is the relevant strong hyperbolicity test for this wave reduction \cite{Sarbach2012}. For a mode $e^{ikx}$, the exact propagator is
\begin{equation}
 \widehat U(\tau,k)=e^{ikv_\pm\tau}
 \begin{pmatrix}1&ik\tau\\0&1\end{pmatrix}\widehat U(0,k).
 \label{eq:exactpropagator}
\end{equation}
Its Euclidean operator norm is
\begin{equation}
 \mathscr A(\chi)=\frac{\sqrt{\chi^2+4}+\chi}{2},
 \qquad \chi=|k|\tau.
 \label{eq:amplification}
\end{equation}
At every fixed positive time, it is unbounded as $|k|\to\infty$. The map between $U$ and $(\partial_xh,\partial_\tau h)$ is a constant invertible transformation at the frozen point. Thus the loss is present in the original tensor wave norm, rather than arising from an added reduction constraint. Explicitly, the admissible tensor data $h(0,x)=0$, $\partial_\tau h(0,x)=e^{ikx}$ have the solution
\begin{equation}
 h(\tau,x)=\tau e^{ik(x+v_\pm\tau)},\qquad
 |\widehat U(\tau,k)|=\sqrt{1+k^2\tau^2}.
 \label{eq:explicitTdata}
\end{equation}
This is a loss of a spatial derivative in the usual wave estimate, not exponential mode growth. The principal equation can still be solved for sufficiently regular data with a stronger norm. No conclusion about a nonlinear strong coupling scale follows from this calculation.

The tensor polarization used here is a physical transverse helicity at radial principal order, rather than a coordinate mode that could be removed by the dipole constraint. Table~\ref{tab:boundaries} summarizes the two boundary results. They concern the quadratic dipole and radial tensor sectors; a full nonlinear degree count is not needed for the stated necessary obstruction.

\end{document}